\documentclass[%
reprint,
superscriptaddress,
amsmath,amssymb,
aps,
prb
floatfix,
showpacs,
]{revtex4-2}

\usepackage{graphicx}
\usepackage{dcolumn}
\usepackage{bm}
\usepackage{hyperref}
\usepackage{arydshln}
\usepackage{ulem}

\usepackage[english]{babel}
\usepackage[utf8]{inputenc}
\usepackage{amssymb,amsfonts,amsmath,mathtools,mathrsfs}
\usepackage{relsize}
\usepackage{mdframed}
\usepackage[margin=0.6in]{geometry}
\usepackage{enumitem}
\usepackage{graphicx}
\usepackage{float}
\usepackage{url,hyperref}
\usepackage[table,usenames,dvipsnames]{xcolor}
\usepackage{gensymb}
\usepackage{grffile}

\hypersetup{
	colorlinks=true,
	linkcolor=Blue,
	filecolor=magenta,      
	urlcolor=Blue,
	citecolor=Blue,
}
\usepackage[bottom]{footmisc}
\usepackage[capitalize]{cleveref}
\usepackage{booktabs}
\usepackage{textcomp}
\usepackage{braket} 

\setcitestyle{line}

\usepackage{amsmath,amssymb,wasysym}

\makeatletter
\def\@fnsymbol#1{\ensuremath{\ifcase#1\or *\or \dagger\or \ddagger\or
   \mathsection\or \mathparagraph\or \|\or **\or \dagger\dagger
   \or \ddagger\ddagger \else\@ctrerr\fi}}
    \makeatother

\begin{document}

\title{Role of composition-tuned disorder on magnetic exchange interactions in NiRuMnGa}

\author{Manikantha Panda}
\address{Department of Physics, National Institute of Technology Andhra Pradesh, Tadepalligudem 534101, India}
\author{Prabuddha Kant Mishra}
\address{Central Advanced Instrumentation Facility, Indian Institute of Science Education and Research Berhampur, Ganjam, Odisha, 760003}
\author{Vinod Kumar}
\address{Laboratoire CRISMAT, CNRS UMR 6508, ENSICAEN, UNICAEN, 6 Boulevard du Maréchal Juin, F-14050 Caen Cedex, France}
\author{Wilfrid Prellier}
\address{Laboratoire CRISMAT, CNRS UMR 6508, ENSICAEN, UNICAEN, 6 Boulevard du Maréchal Juin, F-14050 Caen Cedex, France}
\author{Rahul Sharma}
\address{Department of Physical Sciences, Indian Institute of Science Education and Research Berhampur, Ganjam, Odisha, 760003}
\author{Tapas Paramanik} \email[E-mail: ]{tapas.phys@nitandhra.ac.in}
\address{Department of Physics, National Institute of Technology Andhra Pradesh, Tadepalligudem 534101, India}

\begin{abstract}
We report the experimental investigation of the quaternary Heusler alloy NiRuMnGa, complemented by first-principles calculations. The energetically favorable crystal structure was identified from first-principles calculations and subsequently used as the structural model for refinement of the powder X-ray diffraction data. The refinement shows good agreement with the experimental pattern and reveals substantial B2-type disorder involving Ni/Ru and Mn/Ga site intermixing. The dc magnetic response exhibits pronounced ZFC--FC irreversibility near $\sim48$~K, with the irreversibility temperature shifting toward lower temperatures under increasing magnetic field, indicating a field-sensitive glassy state. The isothermal magnetization remains unsaturated up to 7~T, reaching only $\sim0.4~\mu_{\mathrm{B}}$/f.u., accompanied by pronounced hysteresis, suggesting competing exchange interactions and frustrated magnetic moments. ac susceptibility, together with magnetic relaxation and memory measurements, confirms slow spin dynamics characteristic of a cluster glass state. The lack of a discernible heat-capacity anomaly near the irreversibility temperature further supports the absence of conventional long-range magnetic order. Collectively, these results demonstrate that competition between ferromagnetic and antiferromagnetic interactions in NiRuMnGa leads to magnetic frustration and stabilizes a cluster glass state.

\end{abstract}

\date{\today}


\maketitle

\section{Introduction}

\begin{figure}[http]
\includegraphics[width= 0.8\columnwidth,angle=0,clip=true]{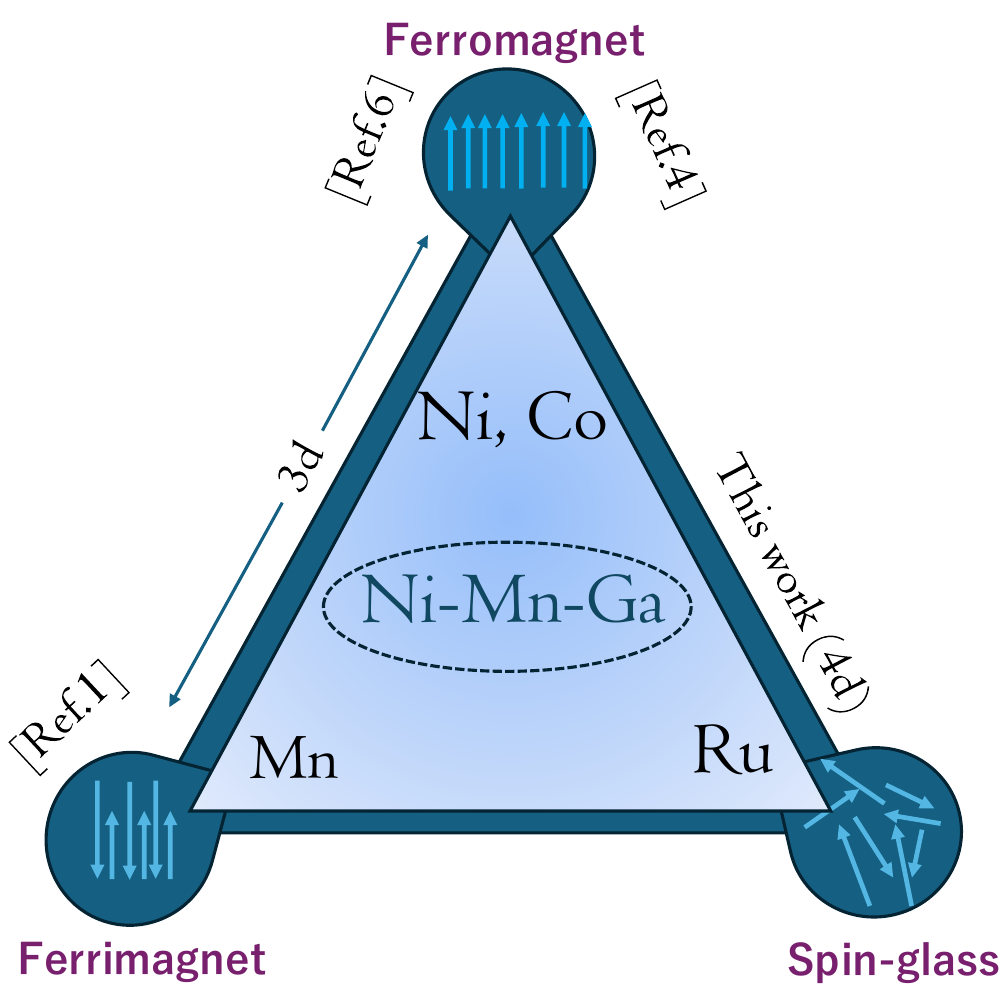}
\caption{(Color online) Compositional tuning of magnetic properties in Ni-Mn-Ga based Heusler alloys.}
\label{Intro}
\end{figure}

The Mn--Ni--Ga alloy system provides a useful platform for exploring the interplay between atomic arrangement, chemical composition, and magnetism, as demonstrated by the distinct magnetic states and disorder effects reported across Ni--Mn--Ga Heusler compositions \cite{Singh2012,DSouza_2014,Antisitedisorder2022,NiCoMnGa}. In these compounds, the magnetic interactions are strongly influenced by the local chemical environment of Mn and the exchange pathways established by the surrounding Ni and Ga (see the schematic Fig.~\ref{Intro}) \cite{Himmetoglu2012,DSouza_2014}. In particular, the balance between superexchange and RKKY-type interactions involving Ni-$d$ and Ga-$p$ states plays an important role in determining their magnetic properties \cite{Himmetoglu2012}. Consequently, changes in composition, site occupancy, or chemical ordering can substantially modify the exchange interactions and magnetic ground state \cite{graf2010heusler}.

This sensitivity is evident in the closely related Ni$_2$MnGa and Mn$_2$NiGa compounds. Ni$_2$MnGa exhibits ferromagnetic order, with its magnetic properties governed by Mn--Mn exchange interactions \cite{NISHIHARA2007191,Himmetoglu2012}, whereas Mn$_2$NiGa adopts an inverse-Heusler structure containing two inequivalent Mn sublattices with antiparallel and unequal moments, resulting in ferrimagnetic order \cite{Singh2012,Singh2014_Mn2NiGa}. Moreover, intrinsic antisite disorder in Mn$_2$NiGa, involving Mn occupation of nominal Ga and Ni sites, has been shown to generate ferromagnetically coupled Mn-rich nanoclusters within the ferrimagnetic matrix \cite{Singh2012,Schaefer2017}. These contrasting behaviors demonstrate that both atomic configuration and site disorder can strongly reshape the local exchange network and produce spatially inhomogeneous magnetism within the Mn--Ni--Ga system.

The ferromagnetic ground state of Ni$_2$MnGa also raises an interesting question regarding the effect of partially replacing Ni with a $4d$ transition-metal element. In Ni$_2$MnGa, ferromagnetic ordering is associated with Mn--Mn exchange interactions, while its critical behavior has been attributed to the presence of long-range RKKY-type interactions \cite{NISHIHARA2007191,Himmetoglu2012,Saleheen2018}. Introducing Ru in place of Ni can modify the local chemical environment and exchange pathways through the involvement of more spatially extended Ru-$4d$ states. Importantly, the closely related NiRuMnSn compound exhibits substantial Ni/Ru site disorder and reentrant spin-glass behavior \cite{NiRuMnSn}. Similar glassy behavior has also been reported in Ru-containing Mn--Ga Heuslers, where disorder and random magnetic correlations have been associated with the emergence of glassy states \cite{CHATTERJEE2021167956,RuMnGa}. Other Ru-containing Heuslers exhibit distinct magnetic ground states, including ferrimagnetic order in Mn$_2$RuGa and antiferromagnetic ordering with low-temperature glassy behavior in FeRuMnSi \cite{Mario_2016,Meinert_2011,PANDA2026187369}. These results demonstrate that Ru-containing Mn-based Heuslers can exhibit diverse magnetic ground states depending on their local chemical environment and degree of disorder.

Against this background, NiRuMnGa provides an interesting system to examine the interplay between chemical disorder and magnetic interactions. The sister compound NiRuMnSn provides an important structural reference, where substantial Ni/Ru disorder is reported, while the Mn and Sn sublattices remain comparatively ordered owing to their large atomic-size difference \cite{NiRuMnSn}. Replacing Sn with Ga is particularly intriguing because the closer atomic sizes of Mn and Ga can facilitate additional site intermixing while preserving the Ni--Ru--Mn framework. Thus, NiRuMnGa offers an opportunity to examine how the main-group element influences chemical disorder and the resulting magnetic exchange network.
In this work, first-principles calculations are combined with powder X-ray diffraction and Rietveld refinement to establish the energetically favorable and experimentally realized structure of NiRuMnGa, followed by comprehensive magnetic and thermodynamic investigations of its ground state. We find substantial B2-type disorder involving both the Ni/Ru and Mn/Ga sublattices, together with competing magnetic interactions evidenced by magnetic measurements. Their coexistence is resulted in magnetic frustration and the emergence of a cluster glass state in NiRuMnGa \cite{panda2026disorder,FeRuMnGa,PANDA2026187369}.

\begin{figure*}
\begin{center}
    \includegraphics[width=1\linewidth]{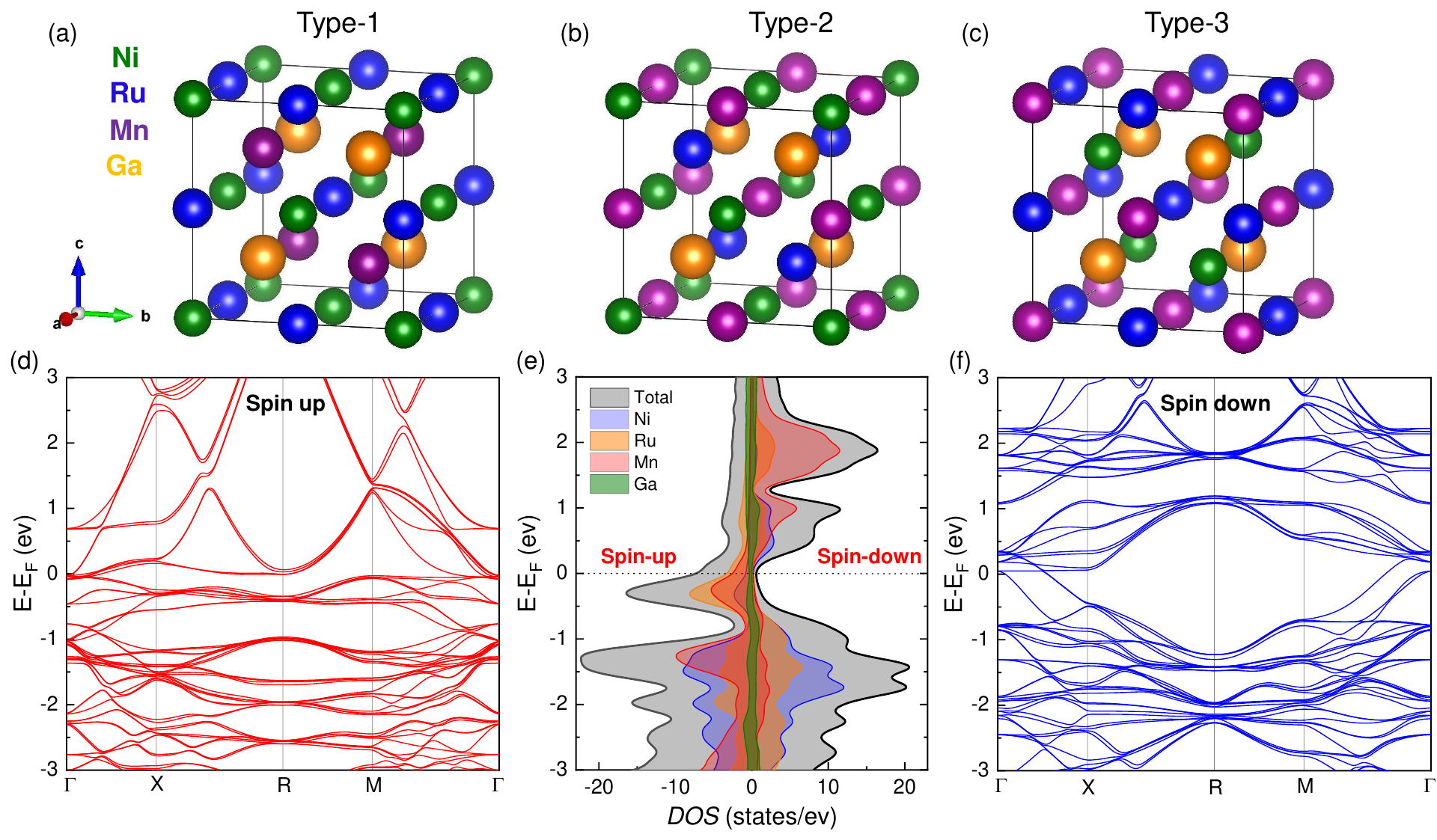}
    \caption{ (Color online) (a-c) Crystal structure of NiRuMnGa considering different Wyckoff position of constituent element while fixing Ga at 4d position. (d) Majority spin-up band. (e) Density of states. (f) Minority spin-down band. }
    \label{cry}
\end{center}
\end{figure*}

\section{Experimental Details}
\begin{figure*}
\begin{center}
    \includegraphics[width=0.8\linewidth]{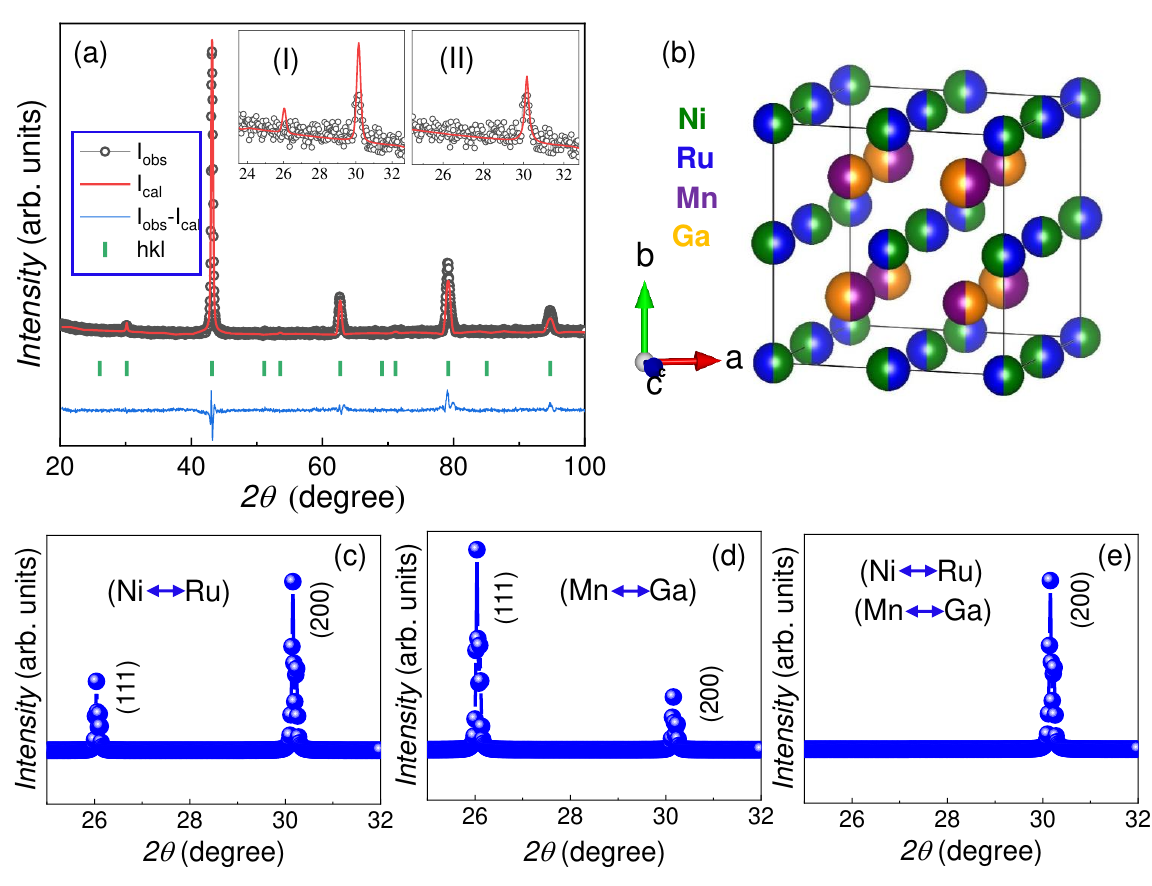}
    \caption{ (Color online) (a) Rietveld refinement of the room-temperature powder X-ray diffraction (PXRD) pattern of polycrystalline NiRuMnGa. The vertical tick marks indicate the Bragg reflection positions, while the blue curve represents the difference between the observed and calculated diffraction profiles. The inset (I) and (II) shows the fitting of super lattice reflections for ordered and disordered structure, respectively. (b) Crystal structure of NiRuMnGa with 50\% B2-type disorder. (c--e) Simulated PXRD patterns showing the enlarged view of the (111) and (200) superlattice reflections generated using VESTA software for (c) 50\% Ni--Ru disorder, (d) 50\% Mn--Ga disorder, and (e) 50\% simultaneous Ni--Ru and Mn--Ga (B2-type) disorder. }
    \label{fig1}
\end{center}
\end{figure*}


Polycrystalline NiRuMnGa was synthesized by the conventional arc-melting technique under a high-purity argon atmosphere. High-purity elemental Ni, Ru, Mn, and Ga ($>$99.9\%, Alfa Aesar, Thermo Scientific) were used as starting materials in their stoichiometric proportions. To compensate for the loss of Mn during melting due to its high vapour pressure, an additional 5\% Mn was added. The alloy button was remelted five times, with the ingot flipped after each melting to ensure compositional homogeneity. The final sample exhibited a negligible weight loss of less than 2\%.
Room-temperature powder X-ray diffraction (PXRD) was carried out using a Malvern Panalytical diffractometer with Cu-$K_{\alpha}$ radiation ($\lambda = 1.5406$~\AA). The obtained diffraction pattern was analyzed by Rietveld refinement using the FULLPROF suite~\cite{FULLPROF}, confirming the formation of a single-phase compound and yielding the crystallographic parameters.
Both dc and ac magnetic measurements were performed using a Quantum Design Magnetic Property Measurement System (MPMS) equipped with a superconducting quantum interference device (SQUID) magnetometer. The dc magnetization measurements were carried out in the temperature range of 2--400~K under magnetic fields up to $\pm$7~T. ac susceptibility measurements were performed between 25 and 75~K at different excitation frequencies and dc field. Heat-capacity measurements were carried out using a Quantum Design Physical Property Measurement System (PPMS).
\section{Computational details}
First-principles calculations were carried out using the plane-wave pseudopotential approach implemented in the Quantum ESPRESSO package~\cite{Giannozzi_2009,Giannozzi_2017} to investigate the structural, electronic, and magnetic properties of the compound. Ultrasoft pseudopotentials were employed for all constituent elements. The exchange--correlation energy was treated within the generalized gradient approximation (GGA) using the Perdew--Burke--Ernzerhof (PBE) functional~\cite{Perdew1996}. A kinetic-energy cutoff of 160~Ry was used for the plane-wave expansion, together with an $6\times6\times6$ Monkhorst--Pack $k$-point mesh for the Brillouin-zone sampling. For the non-self-consistent field (NSCF) calculations, the $k$-point mesh was increased to $16\times16\times16$ to obtain improved sampling and satisfactory convergence. Brillouin-zone integrations were performed using the linear tetrahedron method~\cite{Bloch_1994}. The equilibrium crystal structure was obtained through total-energy minimization, with the atomic positions and lattice parameters relaxed until the residual forces were below $10^{-4}$~Ry/a.u. The convergence criteria for both the total energy and the self-consistent field iterations were set to $10^{-6}$ Ry.

\section{Results and Discussion}

\subsection{Theoretical results}
DFT calculations were performed to investigate the structural properties of NiRuMnGa, considering the LiMgPdSn-type crystal structure. To determine the preferred atomic arrangement, Ga was fixed at the $4d$ Wyckoff position, while Ni, Ru, and Mn were considered at the $4a$, $4b$, and $4c$ sites in three different configurations (Type-I, Type-II, and Type-III), as shown in Fig.~\ref{cry} (a-c). In these configurations, the site sequences are Ni--Ru--Mn, Ni--Mn--Ru, and Mn--Ru--Ni, respectively.
Among the three considered configurations, the Type-I structure is found to be energetically most favorable, as evidenced by its lowest total energy, and is therefore considered the most stable atomic arrangement for NiRuMnGa, with a calculated lattice parameter of 5.92~\AA. This site preference is consistent with the atomic ordering reported for FeRuMnSi \cite{PANDA2026187369} and FeRuCrSi~\cite{ZHANG2023111541}, and can be understood in terms of the valence-electron rule together with the relative atomic-size effects
\cite{Kandpal_2007}.

For this structure, we calculated spin-resolved band structures together with the corresponding DOS, as shown in Fig.~\ref{cry} (d-f). They exhibit a pronounced spin-dependent electronic structure, with distinct differences in the dispersion of the spin-up and spin-down states near the Fermi level. The overall band dispersions indicate the metallic nature of NiRuMnGa, with electronic states present around $E_F$ in both spin channels. The spin-resolved DOS further supports this behavior and reveals a pronounced asymmetry between the two spin channels near the Fermi level.

The calculated spin polarization at the Fermi level, obtained using
\begin{equation}
P(E_F) = \frac{D_\uparrow(E_F)-D_\downarrow(E_F)}
{D_\uparrow(E_F)+D_\downarrow(E_F)} \times 100\%,
\end{equation}
is approximately $85\%$. Although this value is lower than the ideal $100\%$ expected for a perfect half-metal, it indicates a substantial spin polarization of the electronic states around $E_F$. The states near the Fermi level are predominantly derived from the transition-metal $d$ states, with Mn-$3d$ and Ru-$4d$ states providing the major contributions, followed by a comparatively smaller contribution from Ni-$3d$ states. In contrast, the contribution from Ga is negligible in the vicinity of $E_F$. Thus, the low-energy electronic structure is primarily governed by the transition-metal $d$ states, particularly those of Mn and Ru.
The calculated total magnetic moment is $4.4~\mu_{\mathrm{B}}$/f.u., which is in close agreement with the theoretical value of $4~\mu_{\mathrm{B}}$/f.u. expected for NiRuMnGa from the Slater--Pauling (SP) rule \cite{Slater}. The calculated atomic magnetic moments of Ni, Ru, and Mn comes out to be $0.58$, $0.22$, and $3.22~\mu_{\mathrm{B}}$, respectively, suggesting dominant magnetic moment contribution from Mn to the total magnetization, while Ni and Ru provide comparatively smaller contributions. The negligible magnetic contribution from Ga is consistent with its weak DOS contribution near the Fermi level. 
The structural and magnetic properties of NiRuMnGa were subsequently investigated experimentally and are discussed in detail in the following sections.

\subsection{Structural analysis}
Figure~\ref{fig1}(a) presents the room-temperature powder X-ray diffraction (PXRD) pattern (scattered circles) of the powdered polycrystalline NiRuMnGa sample, together with the Rietveld refinement profile (red line), Bragg reflection positions, and the corresponding difference curve. The experimental pattern is well reproduced by the calculated profile for the fundamental Bragg reflections, and no additional reflections attributable to secondary phases are observed, confirming the formation of a single-phase compound within the detection limit of PXRD. The refinement indicates that NiRuMnGa crystallizes in the cubic LiMgPdSn-type structure with space group $F\overline{4}3m$ (No.~216), with a refined lattice parameter of approximately $5.922(3)$~\text{\AA}, matching well with the theoretical obtained value. The atomic configuration was adopted from the theoretically predicted lowest-energy ordered structure, with $4a$ $(0,0,0)$, $4b$ $(\tfrac{1}{2},\tfrac{1}{2},\tfrac{1}{2})$, $4c$ $(\tfrac{1}{4},\tfrac{1}{4},\tfrac{1}{4})$, and $4d$ $(\tfrac{3}{4},\tfrac{3}{4},\tfrac{3}{4})$ Wyckoff positions, respectively.

Although the fundamental reflections are satisfactorily reproduced, noticeable discrepancies are observed for the superlattice reflections, particularly the (111) and (200) reflections. Such discrepancies are commonly encountered in quaternary Heusler alloys and are generally associated with chemical disorder or partial atomic site occupation \cite{panda2026disorder}. Therefore, the degree of chemical ordering was further examined by considering the X-ray scattering factors of the constituent elements. For the present atomic arrangement, the structure factor can be expressed as

\begin{align}
F_{hkl} &= F_{\mathrm{fcc}}\Big[ f_{\mathrm{Ni}}
+ f_{\mathrm{Ru}}e^{\pi i(h+k+l)} \notag\\
&\qquad + f_{\mathrm{Mn}}e^{\frac{\pi i}{2}(h+k+l)}
+ f_{\mathrm{Ga}}e^{-\frac{\pi i}{2}(h+k+l)}\Big],
\end{align}

where $F_{\mathrm{fcc}}=4$ for the allowed fcc reflections and zero otherwise, and $f_{\mathrm{Ni}}$, $f_{\mathrm{Ru}}$, $f_{\mathrm{Mn}}$, and $f_{\mathrm{Ga}}$ represent the corresponding atomic X-ray scattering factors. Accordingly, the structure factors for the (111) and (200) superlattice reflections are given by

\begin{equation}
F_{111}=4\left[(f_{\mathrm{Ni}}-f_{\mathrm{Ru}})+i(f_{\mathrm{Ga}}-f_{\mathrm{Mn}})\right],
\end{equation}

and

\begin{equation}
F_{200}=4\left[(f_{\mathrm{Ni}}+f_{\mathrm{Ru}})-(f_{\mathrm{Mn}}+f_{\mathrm{Ga}})\right].
\end{equation}

These expressions show that the superlattice reflections are highly sensitive to atomic site occupation. Experimentally, the (111) reflection is absent while the (200) reflection remains visible [inset of Fig.~\ref{fig1}(a)], characteristic of B2-type disorder in quaternary Heusler alloys \cite{FeRuMnGa}. To identify the disorder configuration, PXRD patterns were simulated using VESTA \cite{VESTA} for 50\% Ni--Ru and Mn--Ga site interchange separately, as well as their simultaneous interchange [Fig.~\ref{fig1}(c--e)]. The individual Ni--Ru or Mn--Ga mixing models retain the (111) reflection, whereas simultaneous Ni--Ru and Mn--Ga mixing suppresses (111) while preserving (200), in agreement with experiment. The full Rietveld refinement incorporating this site mixing consequently reproduces the experimental PXRD pattern well, indicating significant B2-type disorder involving simultaneous Ni--Ru and Mn--Ga intermixing. Although this analysis establishes the likely nature of the disorder, the individual site occupancies cannot be reliably quantified from laboratory PXRD owing to the comparable X-ray scattering factors of the constituent elements; neutron or synchrotron X-ray diffraction would be required for quantitative determination.


\begin{figure*}[http]
    \centering
    \includegraphics[width=0.8\linewidth]{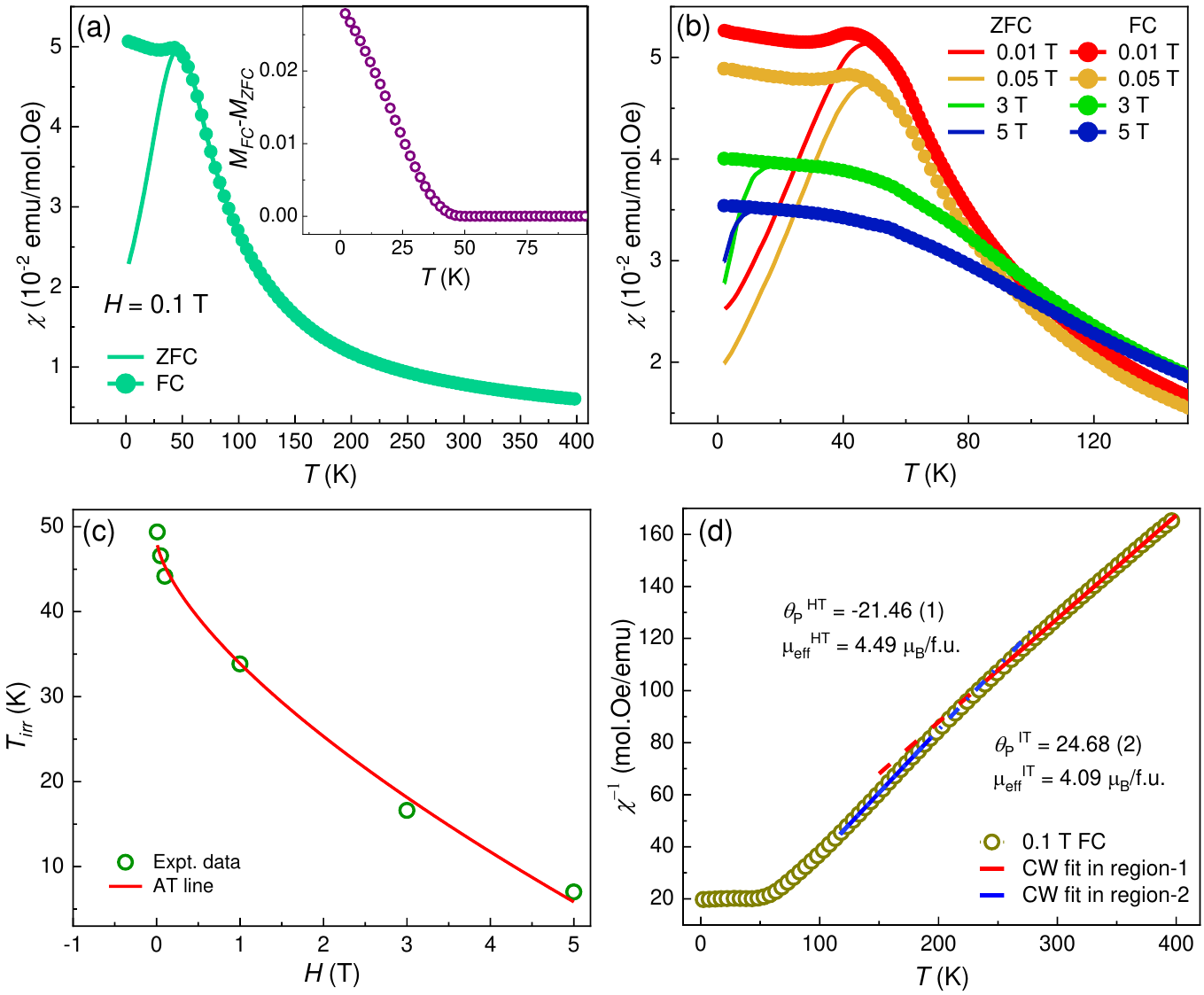}
    \caption{(Color online) (a) Temperature dependence of dc susceptibility $\chi(T)$ measured under zero-field-cooled (ZFC) and field-cooled (FC) protocols at an applied field of 0.1 T for NiRuMnGa. The inset shows $M_{\mathrm{FC}} - M_{\mathrm{ZFC}}$ as a function of temperature. (b) Temperature dependence of magnetization under ZFC and FC conditions at applied fields of 0.01, 0.05, 3 and 5 T. (c) AT line shows the variation of $T_{irr}$ with applied field. (d) Inverse susceptibility $\chi^{-1}(T)$ at an applied field of 0.1 T along with the Curie--Weiss fit using Eq.~(\ref{CW}).}
    \label{MT}
\end{figure*}
\begin{figure*}
    \centering
    \includegraphics[width=0.8\linewidth]{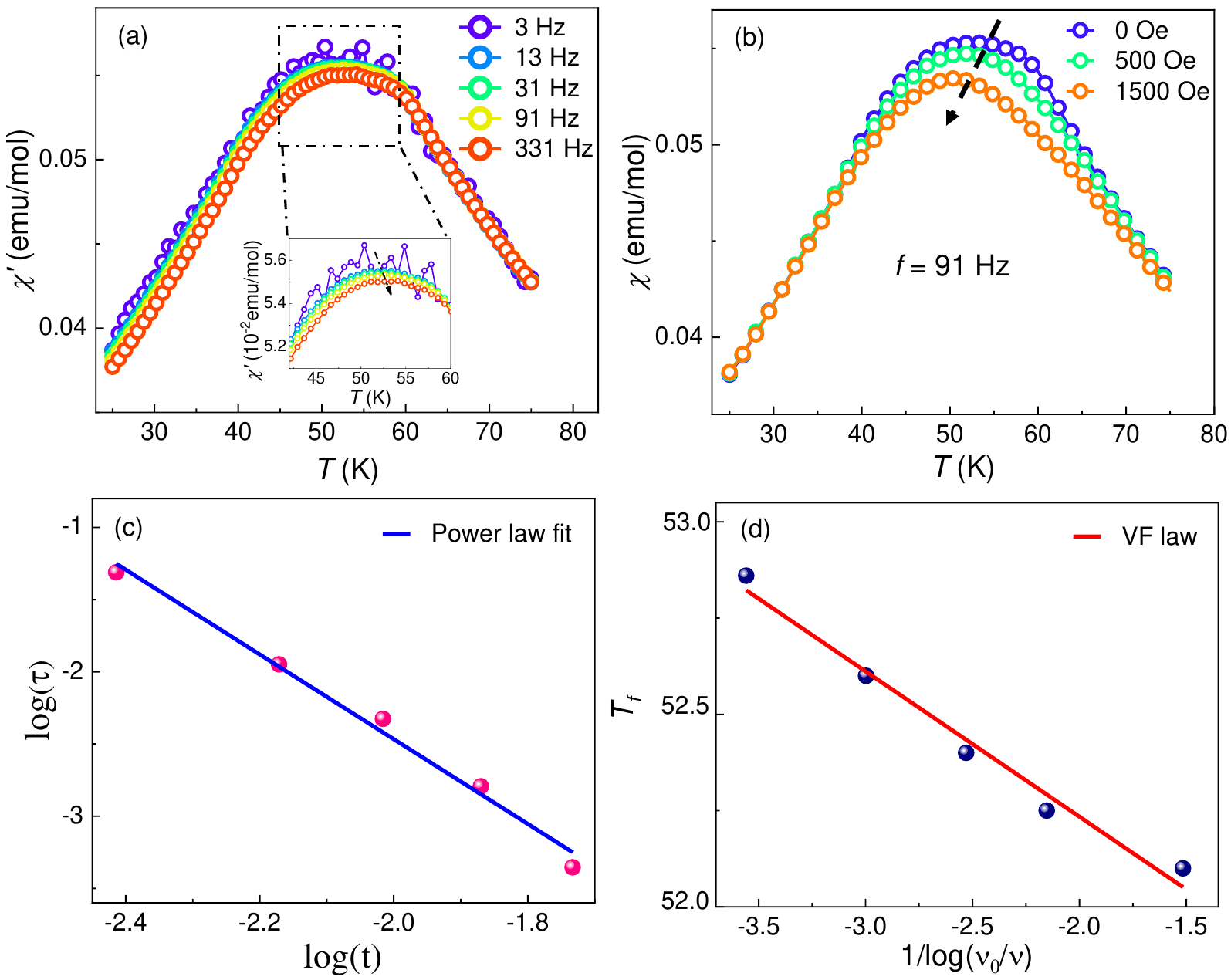}
    \caption{ ac susceptibility data and fitting the shift of ${T_f}$ for NiRuMnGa. (a) Temperature dependence of the real (${\chi'}$) part of the ac magnetic susceptibility measured at $H_{DC}$ = 0 Oe, $H_{AC}$= 0.5 Oe in the temperature range 25 - 75 K. (b) Field dependence of the real (${\chi'}$) part of the ac susceptibility for the frequency 91 Hz. (c) Critical scaling fit using equation \ref{critical}, and (d) Vogel-Fulcher law fit using equation \ref{vogel3}.}
    \label{ac}
\end{figure*}
\subsection{Temperature-dependent magnetic susceptibility}

\hyperref[MT]{Fig.~4(a)} shows the temperature dependence of the dc magnetic susceptibility ($\chi_{\mathrm{dc}}$) of NiRuMnGa measured under zero-field-cooled (ZFC) and field-cooled (FC) protocols in an applied magnetic field of $0.1~\mathrm{T}$ over the temperature range of $2$--$400~\mathrm{K}$. With lowering temperature, the susceptibility increases exhibits a broad maximum at approximately $48~\mathrm{K}$, and for $T<48~K$, ZFC and FC bifurcates, giving rise to magnetic irreversibility. The temperature at which the ZFC and FC curves start to separate is defined as the irreversibility temperature ($T_{\mathrm{irr}}$).
The temperature dependence of $\chi_{\mathrm{dc}}$ was measured under different applied magnetic fields, as presented in \hyperref[MT]{Fig.~4(b)}. It is observed that $T_{\mathrm{irr}}$ shifts progressively towards lower temperatures with increasing magnetic field, indicating field-induced suppression of magnetic irreversibility.
Such field-dependent irreversibility is a characteristic feature of metastable magnetic systems and has been widely reported in spin-glass \cite{Mishra2024}, cluster glass \cite{panda2026disorder, paramanik2015resistivity}, and reentrant magnetic systems \cite{PANDA2026187369}. Similar behavior may also arise from competing ferromagnetic and antiferromagnetic exchange interactions \cite{Mishra2024Te,Lopez2001}, magnetocrystalline anisotropy \cite{Magnetic_anisotropy}, domain-wall pinning \cite{Domainwallpinning}, or chemical disorder \cite{chemical_disorder}, all of which introduce energy barriers for magnetic moment reversal.

To gain further insight into the magnetic irreversibility, the field dependence of $T_{\mathrm{irr}}$ was analyzed using the Almeida--Thouless (AT) model, which describes the phase boundary between the paramagnetic and frozen magnetic states in spin-glass systems. According to the AT model,
\begin{equation}
H = A\left[1-\frac{T_{\mathrm{irr}}(H)}{T_{\mathrm{irr}}(0)}\right]^{3/2},
\end{equation}
where $A$ is a fitting constant and $T_{\mathrm{irr}}(0)$ is the zero-field irreversibility temperature. As shown in \hyperref[MT]{Fig.~4(c)}, the experimental data are well described by the AT relation, indicating a characteristic field-dependent evolution of the magnetic irreversibility consistent with a glassy magnetic state. 


The inverse dc magnetic susceptibility ($\chi^{-1}(T)$) measured under an applied field of $0.1~\mathrm{T}$ was analyzed using the Curie--Weiss law in two different temperature regimes, as shown in Fig.~\ref{MT}(d). The magnetic susceptibility follows the Curie--Weiss relation,

\begin{equation} \label{CW}
\chi=\frac{C}{T-\theta_{\mathrm{P}}},
\end{equation}

where $C$ is the Curie constant and $\theta_{\mathrm{P}}$ is the Weiss temperature. The effective magnetic moment was calculated using

\begin{equation}
\mu_{\mathrm{eff}}=\sqrt{\frac{3k_{\mathrm{B}}C}{N_{\mathrm{A}}}}=\sqrt{8C},
\end{equation}

where $k_{\mathrm{B}}$ is the Boltzmann constant and $N_{\mathrm{A}}$ is Avogadro's number.
The Curie--Weiss fit in the high-temperature region ($250$--$400~\mathrm{K}$) yields an effective magnetic moment of $\mu_{\mathrm{eff}} \approx 4.49~\mu_{\mathrm{B}}$/f.u. and ${\theta_{\mathrm{P}}^{HT}} \approx -21.5~\mathrm{K}$. The negative value of ${\theta_{\mathrm{P}}^{HT}}$ indicates that antiferromagnetic exchange interactions dominate in this temperature range. However, fitting the intermediate-temperature region ($120$--$190~\mathrm{K}$) results in a reduced effective magnetic moment of $\mu_{\mathrm{eff}} \approx 4.03~\mu_{\mathrm{B}}$/f.u. together with a positive ${\theta_{\mathrm{P}}^{IT}}$, suggesting the shift in dominant correlations to ferromagnetic on lowering temperature.
The crossover from a negative to a positive ${\theta_{\mathrm{P}}}$, accompanied by a reduction in the effective magnetic moment, signifies a gradual evolution of the magnetic correlations with decreasing temperature. Such behavior has frequently been reported in magnetically frustrated and glassy systems, generally attributed to the progressive development of short-range magnetic correlations, leading to a crossover from predominantly antiferromagnetic- to ferromagnetic-like interactions \cite{IrMnGa}. 
These observations point towards the presence of competing magnetic interactions \cite{PANDA2026187369}. To understand the microscopic origin of this, we examine it further through ac susceptibility, magnetic relaxation, and memory-effect measurements.



\subsection{ac susceptibility}

Frequency-dependent ac susceptibility measurements were performed across the $T_{irr}$ to probe the magnetic dynamics \cite{Bauer2012,Topping2018}. The $\chi_{\mathrm{ac}}$ was measured from $25$--$75~\mathrm{K}$ at frequencies of $3$--$331~\mathrm{Hz}$ with an ac field of $0.5~\mathrm{Oe}$ under ZFC conditions. As shown in Fig.~\ref{ac}(a), $\chi^{\prime}$ exhibits a peak near $49~\mathrm{K}$, which systematically shifts toward higher temperatures with increasing frequency, accompanied by a slight decrease in peak intensity. This frequency-dependent shift is characteristic of glassy magnetic dynamics and reflects the progressive slowing down of magnetic fluctuations on approaching the frozen state \cite{Pakhira2016}.
The dc-field dependence of $\chi^{\prime}$ was further investigated at 91 Hz, as shown in Fig.~\ref{ac}(b). Increasing the dc field broadens and suppresses the peak while shifting $T_{irr}$ toward lower temperatures. Such field-induced suppression of the irreversibility temperature is consistent with glassy magnetic behavior, where the applied field destabilizes metastable spin configurations and weakens cooperative freezing. Together with the frequency dependence, this behavior provides further evidence for a glassy state arising from competing magnetic interactions \cite{GHOSH2025182719,Mishra2024b}.

To obtain a preliminary assessment of the nature of the glassy magnetic state, the relative shift in the irreversibility temperature with frequency was quantified using the Mydosh parameter, defined as

\begin{equation}
\delta T_f=\frac{\Delta T_f}{T_f\,\Delta\log_{10}f},
\end{equation}

where $\Delta T_f$ is the change in the irreversibility temperature corresponding to the change in excitation frequency $\Delta\log_{10}f$. The Mydosh parameter provides a simple criterion to distinguish between different types of slow magnetic dynamics. Canonical spin-glass systems typically exhibit $\delta T_f$ values in the range of $0.005$--$0.01$, reflecting strong cooperative spin freezing. cluster glass systems generally show intermediate values of $0.01$--$0.08$, arising from the collective freezing of interacting magnetic clusters. In contrast, non-interacting superparamagnetic systems display significantly larger values, usually exceeding $0.10$, owing to the independent thermal blocking of magnetic nanoparticles or clusters \cite{Mydosh1993,Pakhira2016}. For NiRuMnGa, it is found to be 0.006 (calculated between 3 Hz and 331 Hz), placing it near the boundary between canonical spin-glass and cluster glass regimes \cite{Mulder1981,Mulder1982,IrMnGa,Khorwal2022}.

\begin{figure}[http]
\includegraphics[width= 0.9\columnwidth,angle=0,clip=true]{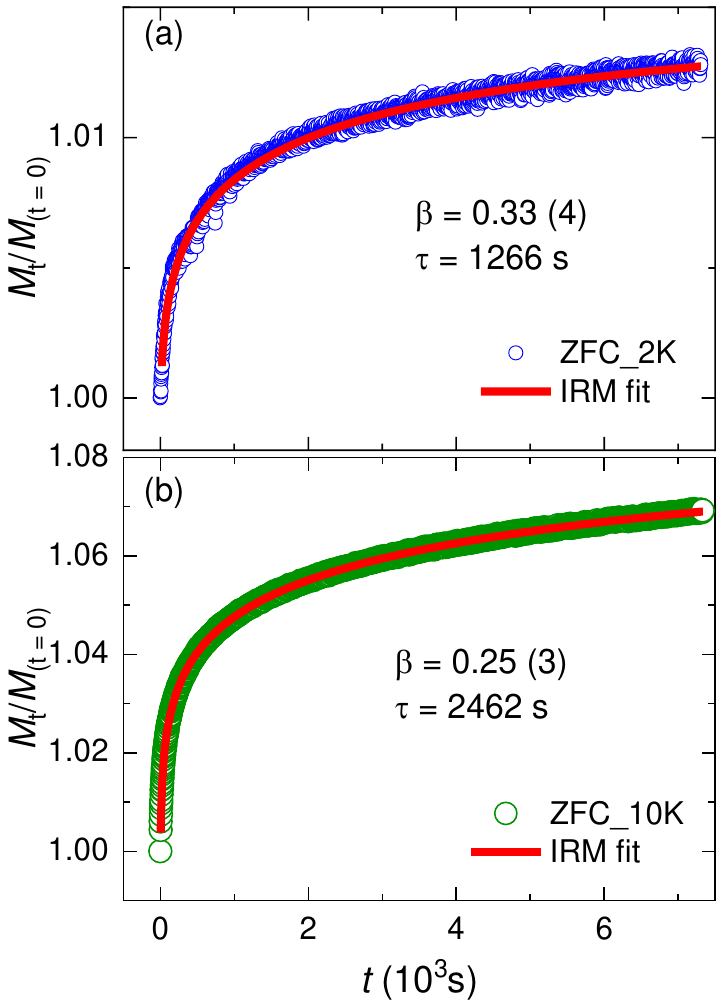}
\caption{(Color online) (a), (b) Time dependence of zero-field-cooled magnetization of NiRuMnGa measured at 2 K and 10 K, respectively, under an applied field of 200 Oe after a wait time 1 hour. The solid lines represent fits to the stretched exponential function [Eq.~(\ref{TRM})].}
\label{TRMfig}
\end{figure}
\begin{figure}
    \centering
    \includegraphics[width=0.9\linewidth]{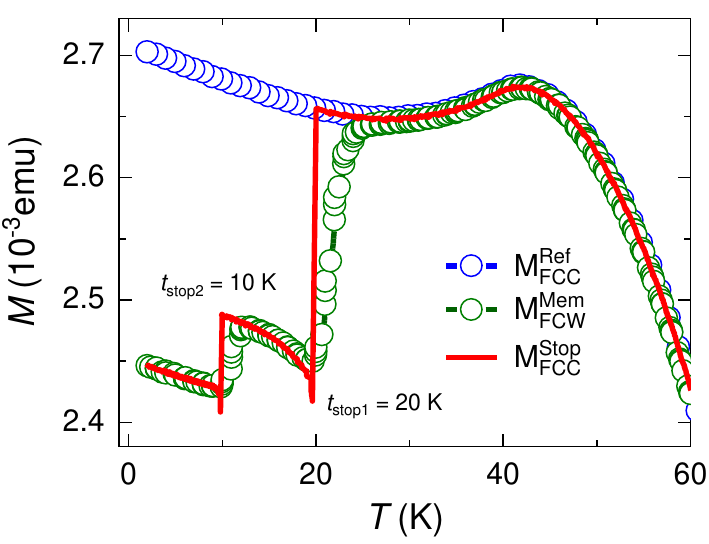}
    \caption{(Color online) Field-cooled (FC) magnetic memory measurement of NiRuMnGa carried out under an applied magnetic field of 100 Oe with intermittent stops at 20 K, and 10 K for a waiting time of 3 h. The interrupted cooling ($M_{\mathrm{FC}}^{\mathrm{stop}}$), reference cooling ($M_{\mathrm{FC}}^{\mathrm{ref}}$), and subsequent warming ($M_{\mathrm{FC}}^{\mathrm{mem}}$) curves demonstrate the recovery of memory at the stop temperatures, characteristic of a cluster glass state.   }
    \label{memory}
\end{figure}

The frequency dependence of the $T_{irr}$ was further analyzed using the critical slowing-down model, which describes the divergence of the spin-relaxation time ($\tau$) as the system approaches the glass transition temperature. The relaxation time follows the dynamic scaling relation,

\begin{equation} \label{critical}
\tau=\tau_0\left(\frac{T_f-T_g}{T_g}\right)^{-z\nu'},
\end{equation}

where $\tau (=1/2\pi f)$ is the characteristic relaxation time corresponding to the measurement frequency $f$, $\tau_0$ is the microscopic spin-flip time, $T_g$ is the glass transition temperature, and $z\nu'$ is the dynamic critical exponent. The solid line in Fig.~\ref{ac}(c) represents the best fit to the experimental data, yielding $z\nu' = 2.93 \pm 0.84$ and $\tau_0 \approx 10^{-8}$ s.
In canonical spin glasses, the microscopic relaxation time is typically of the order of $10^{-10}$--$10^{-13}$ s with a dynamic critical exponent lying in the range of $4$--$12$, reflecting the cooperative freezing of individual atomic spins near the glass transition \cite{Lago2012,Malinowski2011}. In contrast, cluster glass systems generally exhibit significantly larger values of $\tau_0$ ($10^{-7}$--$10^{-10}$ s), since the relaxing entities are not individual spins but interacting clusters with comparatively slower dynamics \cite{Kumar2021,Tapas_Dy5Pd2J}. The obtained value of $\tau_0 \approx 10^{-8}$ s therefore indicates that the magnetic relaxation in NiRuMnGa is governed by the collective dynamics of interacting spin clusters rather than isolated atomic spins.


The frequency-dependent freezing dynamics were further analysed using the Vogel--Fulcher (VF) law, which describes the thermally activated relaxation of interacting magnetic clusters. Unlike the Arrhenius model, which assumes non-interacting magnetic entities, the VF formalism incorporates the effect of inter-cluster interactions that progressively slow down the spin dynamics near the irreversibility temperature. The relaxation time is given by

\begin{equation}\label{vogel1}
\tau=\tau_0\exp\left(\frac{E_a}{k_B(T_f-T_0)}\right),
\end{equation}

where $\tau_0$ is the characteristic microscopic relaxation time, $E_a$ is the activation energy barrier, and $T_0$ is the Vogel--Fulcher temperature, which reflects the strength of inter-cluster interactions. In the absence of such interactions, $T_0 \rightarrow 0$, reducing the VF law to the Arrhenius relation. For linear fitting, Eq.~(\ref{vogel1}) is rewritten as

\begin{equation}\label{vogel3}
T_f=\left(\frac{E_a/k_B}{\ln(\nu_0/\nu)}\right)+T_0,
\end{equation}

where $\nu_0 (=1/\tau_0)$ is the characteristic attempt frequency obtained from the critical slowing-down analysis. The linear fit shown in Fig.~\ref{ac}(d) yields $E_a/k_B = 0.38 \pm 0.03$ K and $T_0 = 51.0 \pm 0.08$ K. 
The activation energy $E_a/k_B$ represents the average energy barrier for the collective reorientation of magnetic clusters, while $T_0$ quantifies the strength of their mutual interactions. A larger $E_a/k_B$ indicates slower relaxation due to higher energy barriers, whereas a larger $T_0$ signifies stronger cooperative interactions that suppress the independent relaxation of magnetic clusters. In the present compound, $T_0 \approx 134(E_a/k_B)$, indicating that the freezing process is governed by strong inter-cluster interactions rather than isolated cluster relaxation \cite{VFlaw}.

Further support for the cluster glass state is provided by the Tholence parameter, $(T_f-T_0)/T_0$, which measures the degree of inter-cluster coupling \cite{Tholence1984}. The obtained value of $\sim0.008$ is consistent with those reported for cluster glass systems \cite{Tholence1984,Bag2018}. Therefore, the Vogel--Fulcher analysis together with the Tholence criterion confirms that the low-temperature magnetic state is characterized by strongly interacting spin clusters, establishing the cluster glass nature of the system.

\subsection{Magnetic relaxation}

To gain further insight into the nonequilibrium magnetic dynamics, magnetic relaxation measurements were carried out at 2 K and 10 K, both below the irreversibility temperature, over a time window of $10^3$ s, as shown in Fig.~\ref{TRMfig}(a,b). Prior to each measurement, the sample was cooled from 300 K to the desired temperature in zero magnetic field and allowed to stabilize for a waiting time ($t_w$) of 1 hour. Subsequently, a small magnetic field of 200 Oe was applied, and the evolution of the magnetization with time was recorded. The relaxation curves were analyzed using the Kohlrausch--Williams--Watts (stretched exponential) function,

\begin{equation}\label{TRM}
M(t)=M_0+M_r\exp\left[-\left(\frac{t}{\tau_r}\right)^{\beta}\right],
\end{equation}

where $M_0$ is the time-independent magnetization, $M_r$ denotes the relaxing magnetic component, $\tau_r$ is the characteristic relaxation time, and $\beta$ ($0<\beta\leq1$) is the stretching exponent. The parameter $\beta$ reflects the distribution of relaxation times in the system: $\beta = 1$ corresponds to a single relaxation process, whereas $\beta < 1$ signifies a broad distribution arising from competing magnetic interactions and a complex energy landscape \cite{Bag2018}.

The best fits yield $\beta = 0.33 \pm 0.008$ and $0.24 \pm 0.001$, with corresponding relaxation times of $\tau_r = 1266$ s and $2462$ s at 2 K and 10 K, respectively. The markedly non-exponential relaxation and the small values of $\beta$ indicate the existence of a wide spectrum of relaxation times, characteristic of interacting spin clusters in a glassy magnetic state. Furthermore, the pronounced temperature dependence of $\tau_r$ suggests thermally activated spin dynamics governed by multiple metastable energy minima. These observations are in good agreement with the behavior reported for glassy systems \cite{FeRuMnGa,Gondh2021, MISHRA2026126189} and provide further evidence that the low-temperature magnetic state of NiRuMnGa is dominated by cooperative relaxation of interacting spin clusters rather than conventional long-range magnetic ordering.

\subsection{Memory effect}
Magnetic memory measurements were performed to further probe the nonequilibrium nature of the frozen magnetic state and is shown Fig.~\ref{memory}, since the ability to retain thermal history is one of the characteristic signatures of glassy magnetic systems. The field-cooled (FC) memory experiment was performed following the protocol proposed by Sun \textit{et al.} \cite{Sun2003}. Initially, the sample was cooled from the paramagnetic region under an applied magnetic field of 500 Oe. During the cooling process, the magnetic field was switched off at the selected halt temperatures of 20~K and~10 K, (all below $T_{irr}$), and the sample was allowed to age for a waiting time of $t_w = 3$ h at each temperature. After each waiting period, the magnetic field was reapplied and the cooling process was resumed until the base temperature of 2~K was reached. The magnetization was measured during this interrupted cooling cycle is denoted as $M_{\mathrm{FC}}^{\mathrm{stop}}$, as shown in Fig.~\ref{memory}. Subsequently, the sample was warmed to the paramagnetic region under the same applied field without any intermediate stops, yielding the memory curve $M_{\mathrm{FC}}^{\mathrm{mem}}$. For comparison, a conventional field-cooled magnetization curve, $M_{\mathrm{FC}}^{\mathrm{ref}}$, was also recorded under identical conditions but without interruption.
As illustrated in Fig.~\ref{memory}, the warming curve $M_{\mathrm{FCW}}^{\mathrm{mem}}$ exhibits distinct anomalies at the respective halt temperatures, closely retracing the interrupted cooling curve. The recovery of these features demonstrates that the system retains information about its previous thermal history despite continuous cooling and subsequent warming. Such memory behavior originates from the slow relaxation of interacting spin clusters trapped in a rugged free-energy landscape containing numerous metastable states. During each waiting period, the magnetic moments partially equilibrate into local energy minima, and these configurations are recovered upon reheating, giving rise to the observed memory dips. The clear manifestation of FC magnetic memory therefore provides strong evidence for nonequilibrium spin dynamics and cooperative freezing of interacting magnetic clusters \cite{Bag2018}, consistent with the cluster glass ground state inferred from the ac susceptibility and magnetic relaxation measurements.

\subsection{Heat capacity}

\begin{figure}[http]
\includegraphics[width= 1\linewidth,angle=0,clip=true]{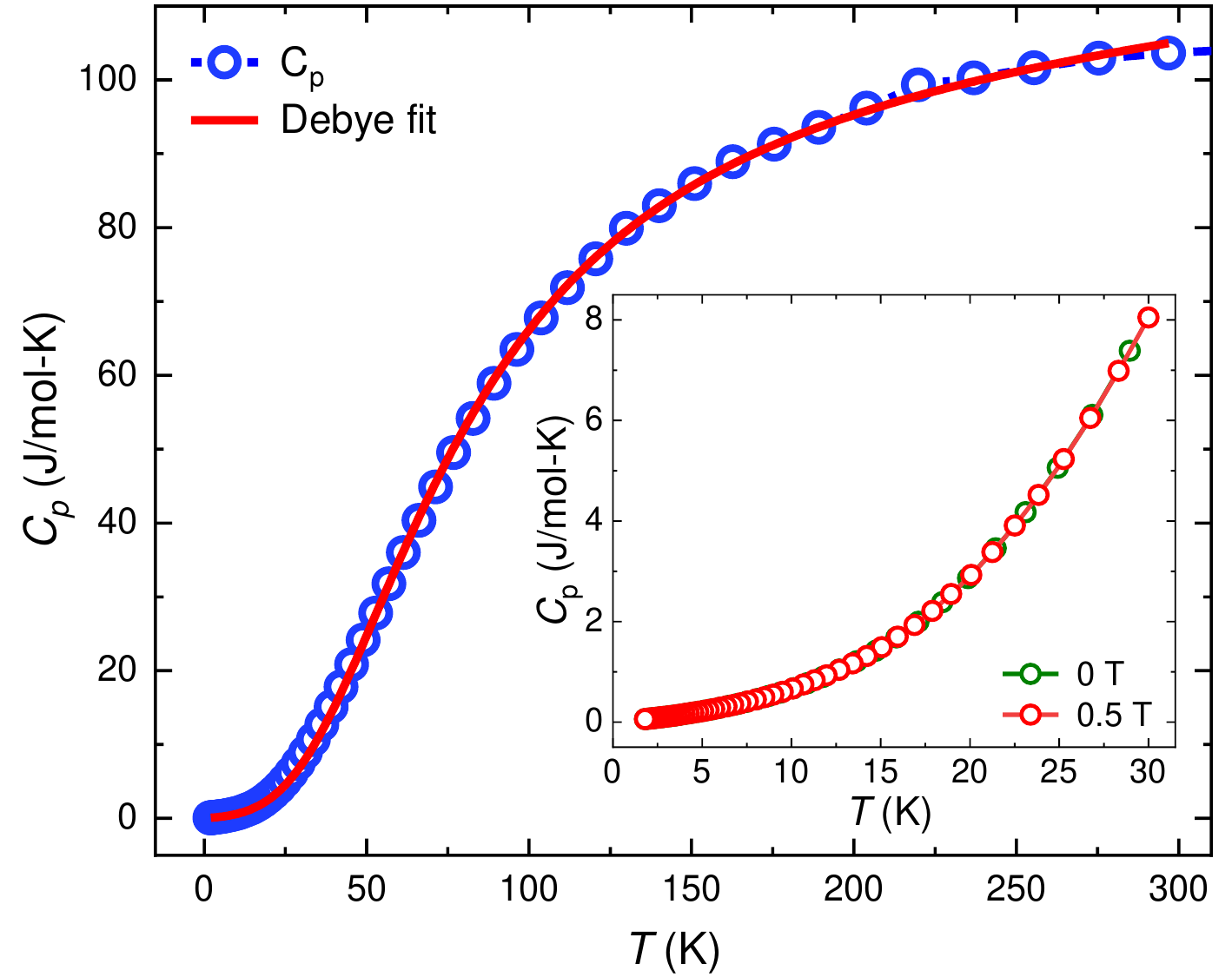}
\caption{(Color online) Heat capacity $C_p(T)$ of NiRuMnGa along with Debye fit as equation \ref{eq:debye}, measured at zero applied field. Inset shows heat capcity at an applied field of 0.5 T. }
\label{HC}
\end{figure}

The temperature dependence of the heat capacity, \(C_p(T)\), was measured over a wide temperature range to examine the thermodynamic signature of magnetic ordering, as shown in Fig.~\ref{HC}. Although the magnetic measurements reveal a freezing transition at $T_{irr} \sim 50$ K, no corresponding $\lambda$-type anomaly or discernible feature is observed in the heat-capacity data. Instead, $C_p(T)$ exhibits a smooth variation throughout the measured temperature range. The absence of a sharp heat-capacity anomaly indicates that the freezing transition is not accompanied by the establishment of conventional long-range magnetic order, rather, it is associated with the gradual development of short-range magnetic correlations and a broad distribution of magnetic entropy, which is characteristic of glassy magnetic systems \cite{Mydosh1993,Binder1986}. Further, field-dependent heat-capacity measurement was performed in the low temperature range 2 - 30 K for applied field of 0.5~T. As shown in the inset of Fig.~\ref{HC}, the heat-capacity curves measured under applied magnetic fields nearly overlap with the zero-field $C(T)$ data, consistent with the freezing of short-range interacting, rather than the development of long-range magnetic order.
At higher temperatures, the heat capacity approaches the Dulong--Petit limit, $C_p = 3nR$, in this regime lattice vibrations dominate the thermal response. The temperature dependence of the heat capacity is well described by the Debye model incorporating an electronic contribution, expressed as
\begin{equation}
C_p(T) = \gamma T + 9nR\left(\frac{T}{\Theta_D}\right)^3
\int_0^{\Theta_D/T}
\frac{x^4 e^x}{(e^x-1)^2}\,dx,
\label{eq:debye}
\end{equation}
where $\gamma$ is the Sommerfeld coefficient, $n$ is the number of atoms per formula unit, $R$ is the universal gas constant, and $\Theta_D$ is the Debye temperature. The fit reproduces the experimental heat-capacity data well and yields a Debye temperature of $\Theta_D \approx 322$~K and a Sommerfeld coefficient of $\gamma = 0.0364$~J\,mol$^{-1}$\,K$^{-2}$. The relatively large value of $\gamma$ compared to FeRuMnSi \cite{PANDA2026187369}, suggests an enhanced electronic density of states at the Fermi level.

\subsection{Isothermal magnetization and exchange bias effect}

\begin{figure}[http]
\includegraphics[width= 0.9\linewidth,angle=0,clip=true]{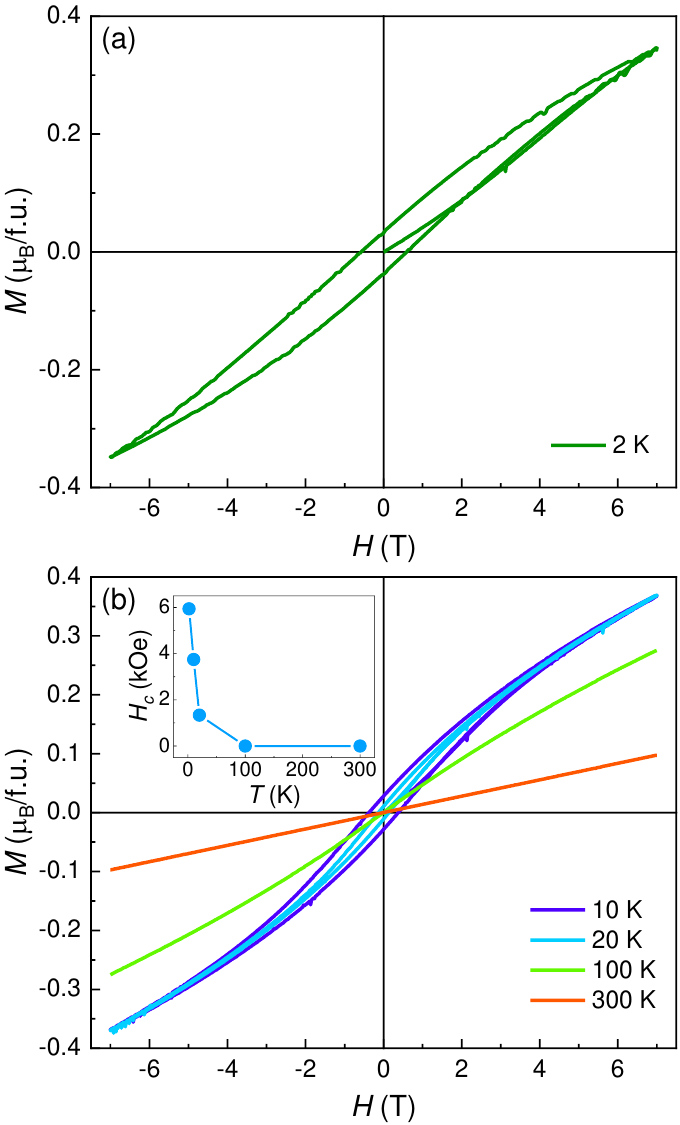}
\caption{(Color online) (a) Isothermal magnetization $M(H)$ of NiRuMnGa measured at 2 K in applied magnetic fields from $-7$ T to $7$ T.   (b)Isothermal magnetization $M(H)$ of NiRuMnGa measured at 10, 20, 100, and 300 K in applied magnetic fields from $-7$ T to $7$ T. (Inset shows the variation of coercivity with respect to the temperature.)}
\label{MH}
\end{figure}

In spin-glass and cluster glass systems, the field dependence of magnetization provides important information on the nature of the magnetic state, magnetic interactions, and field-induced spin rearrangements. To examine these characteristics, isothermal magnetization \(M(H)\) measurements were performed at different temperatures, as shown in Fig.~\ref{MH}. The $M(H)$ curve measured at 2~K exhibits a clear hysteresis with a coercive field ($H_{CO}$) of approximately $0.6$ T, confirming the presence of irreversible magnetization below the irreversibility temperature. The $H_{CO}$ decreases progressively with increasing temperature, as illustrated in the inset of Fig.~\ref{MH}, indicating a gradual weakening of magnetic anisotropy and irreversible spin freezing due to thermal activation \cite{IrMnGa, FeRuMnGa}. Despite the large applied magnetic field of 7~T, the magnetization remains far from saturation and reaches only $\sim0.4~\mu_B$/f.u. This value is substantially lower than the magnetic moment expected from the Slater--Pauling (S--P) rule for the corresponding quaternary Heusler composition \cite{Slater}. The pronounced suppression of the measured moment, together with the unsaturated magnetic response, is inconsistent with a collinear ferromagnetic ground state and instead resembles the behavior of frustrated antiferromagnets, indicating the coexistence of strong antiferromagnetic exchange interactions with ferromagnetic correlations \cite{PANDA2026187369,FeRuMnGa}., consistent with the competing magnetic interactions inferred from the dc susceptibility.

Another remarkable feature of the low-temperature hysteresis loop is that the virgin magnetization curve lies outside the envelope curve over the field range of approximately 2--4.5 T. Such anomalous virgin-curve behavior is frequently observed in magnetically frustrated systems and cluster glasses and is attributed to irreversible field-driven evolution of frozen spin clusters through a rugged free-energy landscape containing numerous metastable states \cite{IrMnSn}. During the initial magnetization process, the frozen clusters progressively overcome anisotropy barriers and evolve towards energetically favorable configurations, whereas subsequent field cycling follows a different reversal path, giving rise to the separation between the virgin and envelope curves. With increasing temperature, the virgin curve gradually merges with the main hysteresis loop as shown in Fig. \ref{MH}(b), reflecting enhanced thermal activation of the spin clusters and the gradual disappearance of the frozen magnetic state.


\begin{figure}[http]
\includegraphics[width= 0.9\linewidth,angle=0,clip=true]{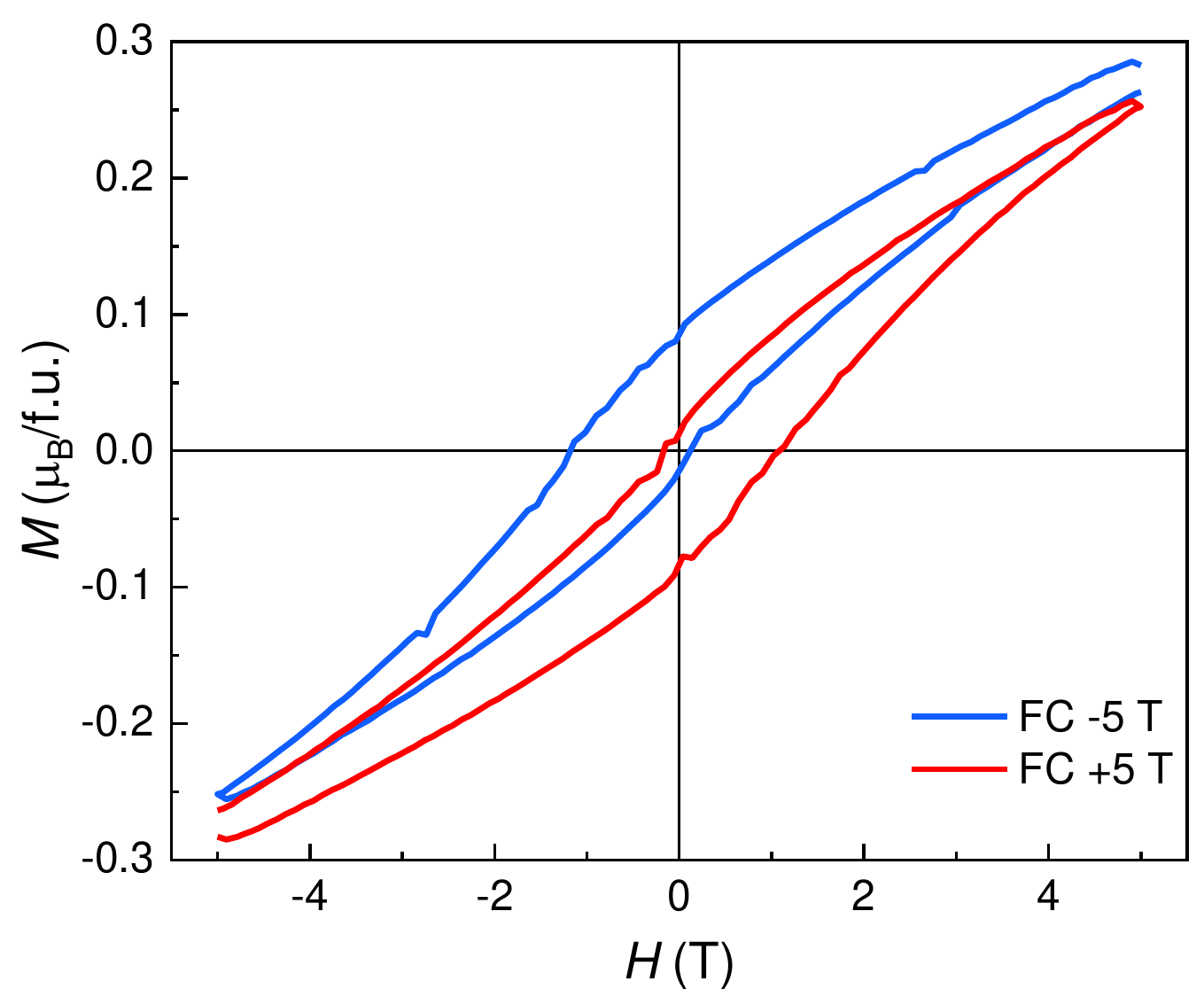}
\caption{(Color online) Magnetic hysteresis loops of NiRuMnGa measured at 2 K after field cooling under $+$5 T and $-$5 T. The direction of the hysteresis-loop displacement reverses with the sign of the cooling field.}
\label{EB}
\end{figure}
A characteristic feature of cluster glass systems is the development of cooling-field-dependent hysteresis-loop shifts, which may arise from exchange bias or the history-dependent response of frozen magnetic moments \cite{Shand2010,Luo2007,Tao2014}. To examine this behavior, hysteresis loops were measured at 2 K after field cooling under \(\pm5\) T. As shown in Fig.~\ref{EB}, the FC loops exhibit a clear displacement with respect to the ZFC loop, and the direction of the shift reverses upon reversal of the cooling-field direction. Such behavior indicates a preferential orientation of the frozen magnetic moments induced by the cooling field.
A shifted hysteresis loop can also result from the minor-loop effect when the magnetization does not reach saturation before field reversal \cite{Coey2010,NOGUES1999203}. In the present case, the FC loops remain unsaturated and are not completely closed even at the maximum applied field, indicating that they represent minor hysteresis loops. Therefore, the observed shift cannot be unambiguously identified as intrinsic exchange bias, which is conventionally established from a fully developed major hysteresis loop. However, the absence of a comparable shift in the ZFC loop suggests that the observed displacement is not solely due to the minor-loop effect. The observed shift is attributed primarily to the history-dependent irreversible response of the frozen interacting spin clusters, consistent with the glassy magnetic state \cite{IrMnGa}.

\section{Conclusion}

The contrasting magnetic states reported in Mn--Ni--Ga Heusler systems highlight the sensitivity of exchange interactions to the local atomic environment, while incorporation of the $4d$ element Ru provides an additional means of tuning this exchange network. In NiRuMnGa, first-principles calculations identify the energetically favorable atomic configuration, while Rietveld refinement reveals substantial B2-type disorder involving Ni/Ru and Mn/Ga site intermixing. The coexistence of this chemical disorder with competing magnetic interactions is reflected in the Curie--Weiss temperatures of $\theta_{\mathrm{P}}\simeq-21.5$~K and $+24.7$~K in the high- and intermediate-temperature regimes, respectively. The resulting complex magnetic state is manifested by ZFC--FC irreversibility near $T_{\mathrm{irr}}\sim48$~K. Frequency-dependent anomalies in the ac susceptibility, together with critical-scaling and Vogel--Fulcher analyses, indicate strongly interacting magnetic clusters with slow dynamics, characterized by $\tau_0 = 10^{-8}$ s, $z\nu = 2.93$ and ${T_0 = 51}$~K, respectively. Relaxation and memory-effect measurements further confirm the nonequilibrium nature of the frozen state. The lack of a discernible heat-capacity anomaly near the irreversibility temperature further supports the absence of conventional long-range magnetic order. Collectively, these results establish significant B2-type chemical disorder and a frustrated cluster glass state in NiRuMnGa. The coexistence of B2-type disorder and competing magnetic interactions highlights the potential of composition-tuned disorder in shaping complex magnetic states in Heusler alloys.

\section*{ACKNOWLEDGMENTS}

M.P. would like to acknowledge NIT Andhra Pradesh for the fellowship. T.P. would like to acknowledge UGC DAE CSR (Grant No. CRS/2021-22/02/487) and SERB (Grant No. CRG/2022/008197) for their financial support. WP and VK thanks to LAFICS supported by CNRS and CEFOPRA (7308-1) project, MP would like to acknowledge the Central Advanced Instrumentation Facility (CAIF), IISER Berhampur, for providing access to the PPMS facility.

\bibliography{NiRuMnGanew}

@article{Bauer2012,
  title = {Magnetic phase diagram of {MnSi} inferred from magnetization and ac susceptibility},
  author = {Bauer, A. and Pfleiderer, C.},
  journal = {Phys. Rev. B},
  volume = {85},
  issue = {21},
  pages = {214418},
  numpages = {16},
  year = {2012},
  month = {Jun},
  publisher = {American Physical Society},
  doi = {10.1103/PhysRevB.85.214418},
  url = {https://link.aps.org/doi/10.1103/PhysRevB.85.214418}
}

@article{Mydosh1993,
  title = {Spin Glasses: An Experimental Introduction},
  journal = {Spin Glasses: An Experimental Introduction},
  author = {Mydosh, J. A.},
  pages = {280},
  numpages = {15},
  year = {1993},
  publisher = {CRC press},
  doi = {https://doi.org/10.1201/9781482295191},

}

@article{Pakhira2016,
  title = {Large magnetic cooling power involving frustrated antiferromagnetic spin-glass state in $\textrm{R}_{2}\textrm{NiSi}_{3}\;(\textrm{R}=\textrm{Gd},\textrm{Er})$},
  author = {Pakhira, Santanu and Mazumdar, Chandan and Ranganathan, R. and Giri, S. and Avdeev, Maxim},
  journal = {Phys. Rev. B},
  volume = {94},
  issue = {10},
  pages = {104414},
  numpages = {15},
  year = {2016},
  month = {Sep},
  publisher = {American Physical Society},
  doi = {10.1103/PhysRevB.94.104414},
  url = {https://link.aps.org/doi/10.1103/PhysRevB.94.104414}
}

@article{Topping2018,
	doi = {10.1088/1361-648x/aaed96},
	url = {https://doi.org/10.1088/1361-648x/aaed96},
	year = {2018},
	month = {nov},
	publisher = {IOP Publishing},
	volume = {31},
	number = {1},
	pages = {013001},
	author = {C V Topping and S J Blundell},
	title = {A.C. susceptibility as a probe of low-frequency magnetic dynamics},
	journal = {J. Phys.: Condens. Matter}
}

@article{Khorwal2022,
  title={Evidence for canonical spin glass behaviour in polycrystalline \textrm{Mn$_{1. 5}$Fe$_{1. 5}$Al Heusler alloy}},
  author={Khorwal, Abhinav Kumar and Dash, Shubhra and Kumar, Arun and Lukoyanov, AV and Shreder, EI and Bitla, Yugandhar and Vasundhara, M and Patra, Ajit K},
  journal={J. Magn. Magn. Mater.},
  volume={546},
  pages={168752},
  year={2022},
  publisher={Elsevier},
  doi = {10.1016/j.jmmm.2021.168752},
  url = {https://doi.org/10.1016/j.jmmm.2021.168752}
}

@article{Tholence1984,
  title={Recent experiments about the spin-glass transition},
  author={Tholence, Jean-Louis},
  journal={Physica B: Condens. Matter},
  volume={126},
  number={1-3},
  pages={157--164},
  year={1984},
  publisher={Elsevier},
  doi = {10.1016/0378-4363(84)90159},
  url = {https://doi.org/10.1016/0378-4363(84)90159-1}
}

@article{Bag2018,
  title = {Cluster spin-glass behavior and memory effect in $\textrm{Cr}_{0.5}\textrm{Fe}_{0.5}\textrm{Ga}$},
  author = {Bag, Pallab and Baral, P. R. and Nath, R.},
  journal = {Phys. Rev. B},
  volume = {98},
  issue = {14},
  pages = {144436},
  numpages = {10},
  year = {2018},
  month = {Oct},
  publisher = {American Physical Society},
  doi = {10.1103/PhysRevB.98.144436},
  url = {https://link.aps.org/doi/10.1103/PhysRevB.98.144436}
}

@article{Gondh2021,
  title = {Large exchange bias and low-temperature glassy state in the frustrated triangular-lattice antiferromagnet $\textrm{Ba}_{3}\textrm{NiIr}_{2}\textrm{O}_{9}$},
  author = {Gondh, Shobha and Patidar, Manju Mishra and Kumar, Kranti and Saravanan, M. P. and Ganesan, V. and Pramanik, A. K.},
  journal = {Phys. Rev. B},
  volume = {104},
  issue = {1},
  pages = {014401},
  numpages = {8},
  year = {2021},
  month = {Jul},
  publisher = {American Physical Society},
  doi = {10.1103/PhysRevB.104.014401},
  url = {https://link.aps.org/doi/10.1103/PhysRevB.104.014401}
}

@article{Shand2010,
  title = {Coercivity and exchange bias of ${\text{Mn}}_{0.25}{\text{Ti}}_{1.1}{\text{S}}_{2}$ in the cluster-glass state},
  author = {Shand, P. M. and Rash, T. and Streicher, M. and Kidd, T. E. and Boyle, K. R. and Strauss, L. H.},
  journal = {Phys. Rev. B},
  volume = {82},
  issue = {21},
  pages = {214413},
  numpages = {8},
  year = {2010},
  month = {Dec},
  publisher = {American Physical Society},
  doi = {10.1103/PhysRevB.82.214413},
  url = {https://link.aps.org/doi/10.1103/PhysRevB.82.214413}
}

@article{Mulder1981,
  title = {Susceptibility of the $\mathrm{Cu}\mathrm{Mn}$ spin-glass: Frequency and field dependences},
  author = {Mulder, C. A. M. and van Duyneveldt, A. J. and Mydosh, J. A.},
  journal = {Phys. Rev. B},
  volume = {23},
  issue = {3},
  pages = {1384--1396},
  numpages = {0},
  year = {1981},
  month = {Feb},
  publisher = {American Physical Society},
  doi = {10.1103/PhysRevB.23.1384},
  url = {https://link.aps.org/doi/10.1103/PhysRevB.23.1384}
}

@article{Mulder1982,
  title = {Frequency and field dependence of the ac susceptibility of the $\mathrm{Au}\mathrm{Mn}$ spin-glass},
  author = {Mulder, C. A. M. and van Duyneveldt, A. J. and Mydosh, J. A.},
  journal = {Phys. Rev. B},
  volume = {25},
  issue = {1},
  pages = {515--518},
  numpages = {0},
  year = {1982},
  month = {Jan},
  publisher = {American Physical Society},
  doi = {10.1103/PhysRevB.25.515},
  url = {https://link.aps.org/doi/10.1103/PhysRevB.25.515}
}

@article{Kumar2021,
  title = {Cluster-glass behavior in the two-dimensional triangular lattice Ising-spin compound $\textrm{Li}_{2}\textrm{Mn}_{3}\textrm{O}_{7}$},
  author = {Kumar, Rahul and Yanda, Premakumar and Sundaresan, A.},
  journal = {Phys. Rev. B},
  volume = {103},
  issue = {21},
  pages = {214427},
  numpages = {10},
  year = {2021},
  month = {Jun},
  publisher = {American Physical Society},
  doi = {10.1103/PhysRevB.103.214427},
  url = {https://link.aps.org/doi/10.1103/PhysRevB.103.214427}
}

@article{Luo2007,
  title = {Spin-glass behavior in hexagonal $\textrm{Na}_{0.70}\textrm{Mn}\textrm{O}_{2}$},
  author = {Luo, L. B. and Zhao, Y. G. and Zhang, G. M. and Guo, S. M. and Li, Z. and Luo, J. L.},
  journal = {Phys. Rev. B},
  volume = {75},
  issue = {12},
  pages = {125115},
  numpages = {7},
  year = {2007},
  month = {Mar},
  publisher = {American Physical Society},
  doi = {10.1103/PhysRevB.75.125115},
  url = {https://link.aps.org/doi/10.1103/PhysRevB.75.125115}
}

@article{Lago2012,
  title = {Three-dimensional Heisenberg spin-glass behavior in \textrm{SrFe${}_{0.90}$Co${}_{0.10}$O${}_{3.0}$}},
  author = {Lago, J. and Blundell, S. J. and Eguia, A. and Jansen, M. and Rojo, T.},
  journal = {Phys. Rev. B},
  volume = {86},
  issue = {6},
  pages = {064412},
  numpages = {8},
  year = {2012},
  month = {Aug},
  publisher = {American Physical Society},
  doi = {10.1103/PhysRevB.86.064412},
  url = {https://link.aps.org/doi/10.1103/PhysRevB.86.064412}
}

@article{Malinowski2011,
  title = {Spin-glass behavior in Ni-doped \textrm{La${}_{1.85}$Sr${}_{0.15}$CuO${}_{4}$}},
  author = {Malinowski, A. and Bezusyy, V. L. and Minikayev, R. and Dziawa, P. and Syryanyy, Y. and Sawicki, M.},
  journal = {Phys. Rev. B},
  volume = {84},
  issue = {2},
  pages = {024409},
  numpages = {17},
  year = {2011},
  month = {Jul},
  publisher = {American Physical Society},
  doi = {10.1103/PhysRevB.84.024409},
  url = {https://link.aps.org/doi/10.1103/PhysRevB.84.024409}
}

@article{Mishra2024,
title = {Experimental studies on noncentrosymmetric {Nd$_{3}$Se$_{4}$}: Strongly correlated noncollinear ferromagnet},
journal = {J. Solid State Chem.},
volume = {334},
pages = {124647},
year = {2024},
issn = {0022-4596},
doi = {https://doi.org/10.1016/j.jssc.2024.124647},
url = {https://www.sciencedirect.com/science/article/pii/S0022459624001014},
author = {Prabuddha Kant Mishra and Ashok Kumar Ganguli},
}

@article{Mishra2024Te,
doi = {10.1088/1402-4896/ad6e2d},
url = {https://dx.doi.org/10.1088/1402-4896/ad6e2d},
year = {2024},
month = {aug},
publisher = {IOP Publishing},
volume = {99},
number = {9},
pages = {095985},
author = {Prabuddha Kant Mishra and Haribrahma Singh and Aarti Gautam and Gaurav Kumar and Rie Umetsu and Ashok Kumar Ganguli},
title = {Observation of competing interactions and field-induced ferromagnetism in noncentrosymmetric cubic {Nd$_{3}$Te$_{4}$} lattice},
journal = {Phys. Scr.}
}

@article{Mishra2024b,
author = {Mishra, Prabuddha Kant and Gautam, Aarti and Singh, Haribrahma and Panda, Soumyakanta and Mohapatra, Niharika and Ganguli, Ashok Kumar},
title = {Interplay of Competing Magnetic Interactions in Noncentrosymmetric {Nd$_{3}$Te$_{4}$} for Enhancing the Magnetocaloric Properties},
journal = {Chem. Mater.},
volume = {36},
number = {12},
pages = {5986-5995},
year = {2024},
doi = {10.1021/acs.chemmater.4c00453},
URL = {https://doi.org/10.1021/acs.chemmater.4c00453}
}

@article{IrMnGa,
  title = {Spin glass behavior in the disordered half-Heusler compound \textrm{IrMnGa}},
  author = {Kroder, Johannes and Manna, Kaustuv and Kriegner, Dominik and Sukhanov, A. S. and Liu, Enke and Borrmann, Horst and Hoser, Andreas and Gooth, Johannes and Schnelle, Walter and Inosov, Dmytro S. and Fecher, Gerhard H. and Felser, Claudia},
  journal = {Phys. Rev. B},
  volume = {99},
  issue = {17},
  pages = {174410},
  numpages = {10},
  year = {2019},
  month = {May},
  publisher = {American Physical Society},
  doi = {10.1103/PhysRevB.99.174410},
  url = {https://link.aps.org/doi/10.1103/PhysRevB.99.174410}
}

@article{FeRuMnGa,
  title = {Experimental observation of spin glass state in the highly disordered quaternary Heusler alloy \textrm{FeRuMnGa}},
  author = {Gupta, Shuvankar and Chakraborty, Sudip and Pakhira, Santanu and Biswas, Anis and Mudryk, Yaroslav and Kumar, Amit and Mukherjee, Bodhoday and Okram, Gunadhor S. and Das, Amitabh and Pecharsky, Vitalij K. and Mazumdar, Chandan},
  journal = {Phys. Rev. B},
  volume = {107},
  issue = {18},
  pages = {184408},
  numpages = {10},
  year = {2023},
  month = {May},
  publisher = {American Physical Society},
  doi = {10.1103/PhysRevB.107.184408},
  url = {https://link.aps.org/doi/10.1103/PhysRevB.107.184408}
}

@article{IrMnSn,
  title = {Carrier localization induced by electron-electron interaction and its impact on the electrical transport and magnetic properties of nonstoichiometric \textrm{IrMnSn}},
  author = {P., Koushik and Mukherjee, K.},
  journal = {Phys. Rev. B},
  volume = {110},
  issue = {17},
  pages = {174419},
  numpages = {12},
  year = {2024},
  month = {Nov},
  publisher = {American Physical Society},
  doi = {10.1103/PhysRevB.110.174419},
  url = {https://link.aps.org/doi/10.1103/PhysRevB.110.174419}
}

@article{NiRuMnSn,
  title = {$4d$ element induced improvement of structural disorder and development of weakly reentrant spin-glass behavior in \textrm{NiRuMnSn}},
  author = {Gupta, Shuvankar and Chakraborty, Sudip and Bhasin, Vidha and Pakhira, Santanu and Biswas, Anis and Mudryk, Yaroslav and Kumar, Amit and Barreteau, Celine and Crivello, Jean-Claude and Das, Amitabh and Jha, S. N. and Bhattacharyya, D. and Pecharsky, Vitalij K. and Alleno, Eric and Mazumdar, Chandan},
  journal = {Phys. Rev. B},
  volume = {108},
  issue = {5},
  pages = {054405},
  numpages = {11},
  year = {2023},
  month = {Aug},
  publisher = {American Physical Society},
  doi = {10.1103/PhysRevB.108.054405},
  url = {https://link.aps.org/doi/10.1103/PhysRevB.108.054405}
}

@article{graf2010heusler,
  title={Heusler compounds—A material class with exceptional properties},
  author={Graf, Tanja and Parkin, Stuart SP and Felser, Claudia},
  journal={IEEE Trans. Magn.},
  volume={47},
  number={2},
  pages={367--373},
  year={2010},
  publisher={IEEE},
doi={10.1109/TMAG.2010.2096229},
url={https://doi.org/10.1109/TMAG.2010.2096229},
}

@article{FULLPROF,
title = {Recent advances in magnetic structure determination by neutron powder diffraction},
journal = {Phys. B: Condens. Matter},
volume = {192},
number = {1},
pages = {55-69},
year = {1993},
issn = {0921-4526},
doi = {https://doi.org/10.1016/0921-4526(93)90108-I},
url = {https://www.sciencedirect.com/science/article/pii/092145269390108I},
author = {Juan Rodríguez-Carvajal}
}

@article{Slater,
  title = {Slater-Pauling behavior and origin of the half-metallicity of the full-Heusler alloys},
  author = {Galanakis, I. and Dederichs, P. H. and Papanikolaou, N.},
  journal = {Phys. Rev. B},
  volume = {66},
  issue = {17},
  pages = {174429},
  numpages = {9},
  year = {2002},
  month = {Nov},
  publisher = {American Physical Society},
  doi = {10.1103/PhysRevB.66.174429},
  url = {https://link.aps.org/doi/10.1103/PhysRevB.66.174429}
}

@article{VFlaw,
title = {The theory of the Vogel-Fulcher law of spin glasses},
journal = {Phys. Lett. A.},
volume = {85},
number = {8},
pages = {467-470},
year = {1981},
issn = {0375-9601},
doi = {https://doi.org/10.1016/0375-9601(81)90441-2},
url = {https://www.sciencedirect.com/science/article/pii/0375960181904412},
author = {S. Shtrikman and E.P. Wohlfarth}
}

@article{Binder1986,
  title = {Spin glasses: Experimental facts, theoretical concepts, and open questions},
  author = {Binder, K. and Young, A. P.},
  journal = {Rev. Mod. Phys.},
  volume = {58},
  issue = {4},
  pages = {801--976},
  numpages = {0},
  year = {1986},
  month = {Oct},
  publisher = {American Physical Society},
  doi = {10.1103/RevModPhys.58.801},
  url = {https://link.aps.org/doi/10.1103/RevModPhys.58.801}
}

@article{Giannozzi_2009,
doi = {10.1088/0953-8984/21/39/395502},
url = {https://dx.doi.org/10.1088/0953-8984/21/39/395502},
year = {2009},
month = {sep},
publisher = {},
volume = {21},
number = {39},
pages = {395502},
author = {Giannozzi, Paolo and Baroni, Stefano and Bonini, Nicola and Calandra, Matteo and Car, Roberto and Cavazzoni, Carlo and Ceresoli, Davide and Chiarotti, Guido L and Cococcioni, Matteo and Dabo, Ismaila and Dal Corso, Andrea and de Gironcoli, Stefano and Fabris, Stefano and Fratesi, Guido and Gebauer, Ralph and Gerstmann, Uwe and Gougoussis, Christos and Kokalj, Anton and Lazzeri, Michele and Martin-Samos, Layla and Marzari, Nicola and Mauri, Francesco and Mazzarello, Riccardo and Paolini, Stefano and Pasquarello, Alfredo and Paulatto, Lorenzo and Sbraccia, Carlo and Scandolo, Sandro and Sclauzero, Gabriele and Seitsonen, Ari P and Smogunov, Alexander and Umari, Paolo and Wentzcovitch, Renata M},
title = {QUANTUM ESPRESSO: a modular and open-source software project for quantum
simulations of materials},
journal = {J. Condens. Matter Phys.}
}

@article{Perdew1996,
  title = {Generalized Gradient Approximation Made Simple},
  author = {Perdew, John P. and Burke, Kieron and Ernzerhof, Matthias},
  journal = {Phys. Rev. Lett.},
  volume = {77},
  issue = {18},
  pages = {3865--3868},
  numpages = {0},
  year = {1996},
  month = {Oct},
  publisher = {American Physical Society},
  doi = {10.1103/PhysRevLett.77.3865},
  url = {https://link.aps.org/doi/10.1103/PhysRevLett.77.3865}
}

@article{Kandpal_2007,
doi = {10.1088/0022-3727/40/6/S01},
url = {https://doi.org/10.1088/0022-3727/40/6/S01},
year = {2007},
month = {mar},
publisher = {},
volume = {40},
number = {6},
pages = {1507},
author = {Kandpal, Hem C and Fecher, Gerhard H and Felser, Claudia},
title = {Calculated electronic and magnetic properties of the half-metallic, transition metal based Heusler compounds},
journal = {J. Phys. D: Appl. Phys.}
}

@article{ZHANG2023111541,
title = {Atomic disorder, magnetic and electronic properties of equiatomic quaternary Heusler alloy {FeRuCrSi}: Experimental and theoretical investigations},
journal = {J. Phys. Chem. Solids},
volume = {181},
pages = {111541},
year = {2023},
issn = {0022-3697},
doi = {https://doi.org/10.1016/j.jpcs.2023.111541},
url = {https://www.sciencedirect.com/science/article/pii/S0022369723003311},
author = {Guangrui Zhang and Hongbo Hao and Yuping Liang and Yu Qiao and Songwei Bai and Heyan Liu and Hongzhi Luo}
}

@article{Tapas_Dy5Pd2J,
author ="Paramanik, Tapas and Samanta, Tapas and Ranganathan, R. and Das, I.",
title  ="Magnetic and magnetocaloric properties of \textrm{Dy$_5$Pd$_2$}: role of magnetic irreversibility",
journal  ="RSC Adv.",
year  ="2015",
volume  ="5",
issue  ="59",
pages  ="47860-47865",
publisher  ="The Royal Society of Chemistry",
doi  ="10.1039/C5RA06970J",
url  ="http://dx.doi.org/10.1039/C5RA06970J"}

@article{GHOSH2025182719,
title = {Role of Fe–Mn interaction in developing cluster glass behaviour and Griffiths phase in ferrimagnetic $\textrm{Gd}(\textrm{Fe}_{0.75}\textrm{Mn}_{0.25})_{2}\textrm{Si}_{2}$},
journal = {J. Alloys Compd.},
volume = {1039},
pages = {182719},
year = {2025},
issn = {0925-8388},
doi = {https://doi.org/10.1016/j.jallcom.2025.182719},
url = {https://www.sciencedirect.com/science/article/pii/S092583882504280X},
author = {Shuvendu Ghosh and Suvayan Saha and Afsar Ahmed and Akshata Magar and Manikantha Panda and Rosni Roy and Rajib Mondal and R. Nath and I. Das and Tapas Paramanik}
}

@article{PANDA2026187369,
title = {Reentrant cluster glass and anomalous transport behavior in disordered quaternary Heusler alloy \textrm{FeRuMnSi}},
journal = {J. Alloys Compd.},
volume = {1061},
pages = {187369},
year = {2026},
issn = {0925-8388},
doi = {https://doi.org/10.1016/j.jallcom.2026.187369},
url = {https://www.sciencedirect.com/science/article/pii/S0925838826014374},
author = {Manikantha Panda and Narayanan T. and Shuvendu Ghosh and Rosni Roy and Akshata Magar and Alapan Bera and Santosh Behara and Soumik Mukhopadhyay and R. Nath and Rajib Mondal and Tapas Paramanik}
}

@article{paramanik2015resistivity,
  title={Resistivity minima in the disordered cluster glass intermetallic compound \textrm{Dy$_5$ Pd$_2$}: influence of quantum interference effects},
  author={Paramanik, Tapas and Das, I},
  journal={RSC Adv.},
  volume={5},
  number={96},
  pages={78406--78413},
  year={2015},
  publisher={Royal Society of Chemistry},
  doi={ 10.1039/C5RA14515E}
}

@article{VESTA,
author = "Momma, Koichi and Izumi, Fujio",
title = "{{\it VESTA}: a three-dimensional visualization system for electronic and structural analysis}",
journal = "J. Appl. Crystallogr.",
year = "2008",
volume = "41",
number = "3",
pages = "653--658",
month = "Jun",
doi = {10.1107/S0021889808012016},
url = {https://doi.org/10.1107/S0021889808012016},
}

@article{Giannozzi_2017,
doi = {10.1088/1361-648X/aa8f79},
url = {https://doi.org/10.1088/1361-648X/aa8f79},
year = {2017},
month = {oct},
publisher = {IOP Publishing},
volume = {29},
number = {46},
pages = {465901},
author = {Giannozzi, P and Andreussi, O and Brumme, T and Bunau, O and Buongiorno Nardelli, M and Calandra, M and Car, R and Cavazzoni, C and Ceresoli, D and Cococcioni, M and Colonna, N and Carnimeo, I and Dal Corso, A and de Gironcoli, S and Delugas, P and DiStasio, R A and Ferretti, A and Floris, A and Fratesi, G and Fugallo, G and Gebauer, R and Gerstmann, U and Giustino, F and Gorni, T and Jia, J and Kawamura, M and Ko, H-Y and Kokalj, A and Küçükbenli, E and Lazzeri, M and Marsili, M and Marzari, N and Mauri, F and Nguyen, N L and Nguyen, H-V and Otero-de-la-Roza, A and Paulatto, L and Poncé, S and Rocca, D and Sabatini, R and Santra, B and Schlipf, M and Seitsonen, A P and Smogunov, A and Timrov, I and Thonhauser, T and Umari, P and Vast, N and Wu, X and Baroni, S},
title = {Advanced capabilities for materials modelling with Quantum ESPRESSO},
journal = {J. Phys.: Condens. Matter}
}

@article{Bloch_1994,
  title = {Improved tetrahedron method for Brillouin-zone integrations},
  author = {Bl\"ochl, Peter E. and Jepsen, O. and Andersen, O. K.},
  journal = {Phys. Rev. B},
  volume = {49},
  issue = {23},
  pages = {16223--16233},
  numpages = {0},
  year = {1994},
  month = {Jun},
  publisher = {American Physical Society},
  doi = {10.1103/PhysRevB.49.16223},
  url = {https://link.aps.org/doi/10.1103/PhysRevB.49.16223}
}

@article{NISHIHARA2007191,
title = {Magnetization processes near the Curie temperatures of the itinerant ferromagnets, {Ni$_{2}$MnGa} and pure nickel},
journal = {J. Alloys Compd.},
volume = {442},
number = {1},
pages = {191-193},
year = {2007},
note = {Proceedings of the 15th International Conference on Solid Compounds of Transition Elements},
issn = {0925-8388},
doi = {https://doi.org/10.1016/j.jallcom.2006.09.154},
url = {https://www.sciencedirect.com/science/article/pii/S0925838807002721},
author = {H. Nishihara and K. Komiyama and I. Oguro and T. Kanomata and V. Chernenko}
}

@article{Singh2012,
  author  = {Singh, Sanjay and Rawat, R. and Muthu, S. Esakki and D'Souza, S. W. and Suard, E. and Senyshyn, A. and Banik, S. and Rajput, P. and Bhardwaj, S. and Awasthi, A. M.},
  title   = {{Spin-valve-like magnetoresistance in Mn$_2$NiGa at room temperature}},
  journal = {Phys. Rev. Lett.},
  volume  = {109},
  pages   = {246601},
  year    = {2012},
  doi     = {10.1103/PhysRevLett.109.246601}
}

@article{DSouza_2014,
doi = {10.1088/0953-8984/26/50/506001},
url = {https://doi.org/10.1088/0953-8984/26/50/506001},
year = {2014},
month = {nov},
publisher = {IOP Publishing},
volume = {26},
number = {50},
pages = {506001},
author = {D'Souza, Sunil Wilfred and Roy, Tufan and Barman, Sudipta Roy and Chakrabarti, Aparna},
title = {{Magnetic properties and electronic structure of Mn–Ni–Ga magnetic shape memory alloys}},
journal = {J. Phys.: Condens. Matter}
}

@article{Antisitedisorder2022,
  title   = {{Role of antisite disorder in the martensitic transition of Ni$_{2-x}$Mn$_{1+x}$Ga}},
  journal = {Intermetallics},
  volume  = {145},
  pages   = {107613},
  year    = {2022},
  doi     = {10.1016/j.intermet.2022.107613}
}

@article{Himmetoglu2012,
  author  = {Himmetoglu, B. and Wentzcovitch, R. M. and Cococcioni, M.},
  title   = {{First-principles study of the electronic structure and magnetic properties of Ni$_2$MnGa}},
  journal = {Phys. Rev. B},
  volume  = {86},
  pages   = {144413},
  year    = {2012}
}

@article{Singh2014_Mn2NiGa,
  author  = {Singh, Sanjay and Muthu, S. Esakki and Senyshyn, Anatoliy and Rajput, Parasmani and others},
  title   = {{Inverse magnetocaloric effect in Mn$_2$NiGa and Mn$_{1.75}$Ni$_{1.25}$Ga magnetic shape memory alloys}},
  journal = {Appl. Phys. Lett.},
  volume  = {104},
  number  = {5},
  pages   = {051905},
  year    = {2014},
  doi     = {10.1063/1.4863742}
}

@article{Mario_2016,
  title = {Designing a fully compensated half-metallic ferrimagnet},
  author = {\ifmmode \check{Z}\else \v{Z}\fi{}ic, Mario and Rode, Karsten and Thiyagarajah, Naganivetha and Lau, Yong-Chang and Betto, Davide and Coey, J. M. D. and Sanvito, Stefano and O'Shea, Kerry J. and Ferguson, Ciaran A. and MacLaren, Donald A. and Archer, Thomas},
  journal = {Phys. Rev. B},
  volume = {93},
  issue = {14},
  pages = {140202(R)},
  numpages = {5},
  year = {2016},
  month = {Apr},
  publisher = {American Physical Society},
  doi = {10.1103/PhysRevB.93.140202},
  url = {https://link.aps.org/doi/10.1103/PhysRevB.93.140202}
}

@article{Meinert_2011,
doi = {10.1088/0953-8984/23/11/116005},
url = {https://doi.org/10.1088/0953-8984/23/11/116005},
year = {2011},
month = {mar},
publisher = {},
volume = {23},
number = {11},
pages = {116005},
author = {Meinert, Markus and Schmalhorst, Jan-Michael and Reiss, Günter},
title = {Exchange interactions and Curie temperatures of
Mn2CoZ 
compounds},
journal = {J. Phys.: Condens. Matter}
}

@article{CHATTERJEE2021167956,
title = {Emergence of compensated ferrimagnetic state in {Mn$_{2-x}$Ru$_{1+x}$Ga} (x = 0.2, 0.5) alloys},
journal = {J. Magn. Magn. Mater.},
volume = {532},
pages = {167956},
year = {2021},
issn = {0304-8853},
doi = {https://doi.org/10.1016/j.jmmm.2021.167956},
url = {https://www.sciencedirect.com/science/article/pii/S0304885321002328},
author = {Snehashish Chatterjee and Prabir Dutta and Pintu Singha and Saurav Giri and Aritra Banerjee and Subham Majumdar}
}

@article{RuMnGa,
  title = {Origin of magnetic ordering in half-Heusler {RuMnGa}},
  author = {Chakraborty, Sudip and Gupta, Shuvankar and Pakhira, Santanu and Biswas, Anis and Mudryk, Yaroslav and Choudhary, Renu and Kumar, Amit and Das, Amitabh and Mazumdar, Chandan},
  journal = {Phys. Rev. B},
  volume = {108},
  issue = {5},
  pages = {054430},
  numpages = {12},
  year = {2023},
  month = {Aug},
  publisher = {American Physical Society},
  doi = {10.1103/PhysRevB.108.054430},
  url = {https://link.aps.org/doi/10.1103/PhysRevB.108.054430}
}

@article{Schaefer2017,
  author  = {Schaefer, D. M. and de Oliveira, A. J. A. and de Camargo, P. C.
             and Varalda, J. and Schreiner, W. H. and Mosca, D. H.},
  title   = {{Exchange-bias reversal in Mn$_{2-x}$Ni$_{1+x}$Ga films with antisite disorder}},
  journal = {Intermetallics},
  volume  = {91},
  pages   = {22--30},
  year    = {2017},
  doi     = {10.1016/j.intermet.2017.08.001}
}

@misc{panda2026disorder,
      title={Disorder-driven competing magnetic interactions and glassy magnetic behavior in quaternary Heusler alloy {FeRuMnGe}}, 
      author={Manikantha Panda and Prabuddha Kant Mishra and Sonali S. Pradhan and Aarti Gautam and Bhagyashree Pol and P D Babu and Ashok Kumar Ganguli and V. Kanchana and Tapas Paramanik},
      year={2026},
      eprint={2609.02506},
      archivePrefix={arXiv},
      primaryClass={cond-mat.str-el},
      url={https://arxiv.org/abs/2609.02506}, 
}

@article{MISHRA2026126189,
title = {Signatures of competing magnetic interactions and correlated transport in noncentrosymmetric {Ce$_{3}$Se$_{4}$}},
journal = {J. Solid State Chem.},
volume = {362},
pages = {126189},
year = {2026},
issn = {0022-4596},
doi = {https://doi.org/10.1016/j.jssc.2026.126189},
url = {https://www.sciencedirect.com/science/article/pii/S0022459626003853},
author = {Prabuddha Kant Mishra and Moumita Naskar and Aarti Gautam and Haribrahma Singh and Rie Umetsu and Ashok Kumar Ganguli}
}

@article{Lopez2001,
  title = {Magnetic relaxation behavior in ${\mathrm{La}}_{0.5}{\mathrm{Ca}}_{0.5}{\mathrm{MnO}}_{3}$ and ${\mathrm{Nd}}_{0.5}{\mathrm{Sr}}_{0.5}{\mathrm{MnO}}_{3}$},
  author = {L\'opez, J. and Lisboa-Filho, P. N. and Passos, W. A. C. and Ortiz, W. A. and Araujo-Moreira, F. M. and de Lima, O. F. and Schaniel, D. and Ghosh, K.},
  journal = {Phys. Rev. B},
  volume = {63},
  issue = {22},
  pages = {224422},
  numpages = {9},
  year = {2001},
  month = {May},
  publisher = {American Physical Society},
  doi = {10.1103/PhysRevB.63.224422},
  url = {https://link.aps.org/doi/10.1103/PhysRevB.63.224422}
}

@article{Magnetic_anisotropy,
title = {Magnetic after-effect in Ni–Mn–Sb Heusler alloy},
journal = {J. Magn. Magn. Mater.},
volume = {320},
number = {5},
pages = {617-621},
year = {2008},
issn = {0304-8853},
doi = {https://doi.org/10.1016/j.jmmm.2007.07.023},
url = {https://www.sciencedirect.com/science/article/pii/S0304885307008268},
author = {S. Chatterjee and S. Giri and S. Majumdar and S.K. De}
}

@article{Domainwallpinning,
  title = {Temperature dependence of exchange bias in polycrystalline ferromagnet-antiferromagnet bilayers},
  author = {Stiles, M. D. and McMichael, R. D.},
  journal = {Phys. Rev. B},
  volume = {60},
  issue = {18},
  pages = {12950--12956},
  numpages = {0},
  year = {1999},
  month = {Nov},
  publisher = {American Physical Society},
  doi = {10.1103/PhysRevB.60.12950},
  url = {https://link.aps.org/doi/10.1103/PhysRevB.60.12950}
}

@article{chemical_disorder,
title = {Memory effect in spin-chain single crystal {Ca$_{3}$Co$_{1.62}$Mn$_{0.38}$O$_{6}$}},
journal = {J. Alloys Compd.},
volume = {620},
pages = {97-100},
year = {2015},
issn = {0925-8388},
doi = {https://doi.org/10.1016/j.jallcom.2014.09.119},
url = {https://www.sciencedirect.com/science/article/pii/S0925838814022920},
author = {M.Y. Ruan and Z.W. Ouyang and Y.M. Guo and Y.C. Sun and J.J. Cheng and Z.C. Xia and G.H. Rao}
}

@article{Sun2003,
  title = {Memory Effects in an Interacting Magnetic Nanoparticle System},
  author = {Sun, Young and Salamon, M. B. and Garnier, K. and Averback, R. S.},
  journal = {Phys. Rev. Lett.},
  volume = {91},
  issue = {16},
  pages = {167206},
  numpages = {4},
  year = {2003},
  month = {Oct},
  publisher = {American Physical Society},
  doi = {10.1103/PhysRevLett.91.167206},
  url = {https://link.aps.org/doi/10.1103/PhysRevLett.91.167206}
}

@article{Tao2014,
  author  = {Tao, Q. Z. and Hu, C. F. and Lin, S. and Zhang, H. B. and
             Li, F. Z. and Qu, D. and Wu, M. L. and Sun, Y. P. and
             Sakka, Y. and Barsoum, M. W.},
  title   = {Coexistence of Ferromagnetic and a Re-entrant Cluster Glass State
             in the Layered Quaternary ({Cr}$_{1-x}${Mn}$_x$)$_2${GeC}},
  journal = {Mater. Res. Lett.},
  volume  = {2},
  number  = {4},
  pages   = {192--198},
  year    = {2014},
  doi     = {10.1080/21663831.2014.909542}
}

@book{Coey2010,
  author    = {Coey, J. M. D.},
  title     = {Magnetism and Magnetic Materials},
  publisher = {Cambridge University Press},
  address   = {Cambridge, UK},
  year      = {2010},
  isbn      = {9780521816144},
  doi       = {10.1017/CBO9780511845000}
}

@article{NOGUES1999203,
title = {Exchange bias},
journal = {J. Magn. Magn. Mater.},
volume = {192},
number = {2},
pages = {203-232},
year = {1999},
issn = {0304-8853},
doi = {https://doi.org/10.1016/S0304-8853(98)00266-2},
url = {https://www.sciencedirect.com/science/article/pii/S0304885398002662},
author = {J Nogués and Ivan K Schuller}
}

@article{NiCoMnGa,
  title = {Ordering tendencies and electronic properties in quaternary Heusler derivatives},
  author = {Neibecker, Pascal and Gruner, Markus E. and Xu, Xiao and Kainuma, Ryosuke and Petry, Winfried and Pentcheva, Rossitza and Leitner, Michael},
  journal = {Phys. Rev. B},
  volume = {96},
  issue = {16},
  pages = {165131},
  numpages = {14},
  year = {2017},
  month = {Oct},
  publisher = {American Physical Society},
  doi = {10.1103/PhysRevB.96.165131},
  url = {https://link.aps.org/doi/10.1103/PhysRevB.96.165131}
}

@article{Saleheen2018,
    author = {Us Saleheen, Ahmad and Chen, Jing-Han and Young, David P. and Dubenko, Igor and Ali, Naushad and Stadler, Shane},
    title = {Critical behavior in {Ni$_{2}$MnGa} and {Ni$_{2}$Mn$_{0.85}$Cu$_{0.15}$Ga}},
    journal = {J. Appl. Phys.},
    volume = {123},
    number = {20},
    pages = {203904},
    year = {2018},
    month = {05},
    issn = {0021-8979},
    doi = {10.1063/1.5025196},
    url = {https://doi.org/10.1063/1.5025196},
}

\end{document}